\documentclass[
 reprint,
 amsmath,amssymb,
 aps,
]{revtex4-1}

\usepackage{subcaption}
\usepackage{dcolumn}
\usepackage{bm}
\usepackage{amsmath}
\usepackage{physics}
\usepackage{multirow}
\usepackage{scrextend}
\usepackage{graphicx}
\usepackage[T1]{fontenc}
\usepackage{algorithm}
\usepackage{algorithmicx}
\usepackage{algpseudocode}
\usepackage{blkarray}

\begin{document}

\preprint{APS/123-QED}

\title{Quantum link bootstrapping using a RuleSet-based communication protocol}

\author{Takaaki Matsuo}
\altaffiliation{Keio University Shonan Fujisawa Campus, 5322 Endo, Fujisawa, Kanagawa 252-0882, Japan}
\author{Cl\'ement Durand}
\altaffiliation{\'Ecole polytechnique, France}
\author{Rodney Van Meter}
\altaffiliation{Keio University Shonan Fujisawa Campus, 5322 Endo, Fujisawa, Kanagawa 252-0882, Japan}

\date{\today}

\begin{abstract}

Establishing end-to-end quantum connections requires quantified link characteristics, and operations need to coordinate decision-making between nodes across a network.
We introduce the RuleSet-based communication protocol for supporting quantum operations over distant nodes to minimize classical packet transmissions for guaranteeing synchronicity.
RuleSets are distributed to nodes along a path at connection set up time, and hold lists of operations that need to be performed in real time.
 We simulate the RuleSet-based quantum link bootstrapping protocol, which consists of recurrent purifications and link-level tomography,
  to quantify the quantum link fidelity and its throughput.
Our Markov-Chain Monte-Carlo simulation includes various error sources, such as the memory error, gate error and channel error, modeled on currently available hardware.
We found that when two quantum nodes, each with 100 memory qubits capable of emitting photons ideally to the optical fiber, are physically connected with a 10km MeetInTheMiddle link, the Recurrent Single selection - Single error purification (RSs-Sp) protocol is capable of bringing up the fidelity from an average input $F_{r}=0.675$ to around $F_{r}=0.865$ with a generation rate of 1106 Bell pairs per second, as determined by simulated tomography.
The system gets noisier with longer channels, in which case errors may develop faster than the purification gain.
In such a situation, a stronger purification method, such as the double selection-based purification, shows an advantage for improving the fidelity.
The knowledge acquired from bootstrapping can later be distributed to nodes within the same network, and used for other purposes such as route selection.

\end{abstract}

\maketitle

\section{\label{sec:intro}INTRODUCTION}
\begin{figure}[htbp]
  \center
  \includegraphics[keepaspectratio,scale=0.33]{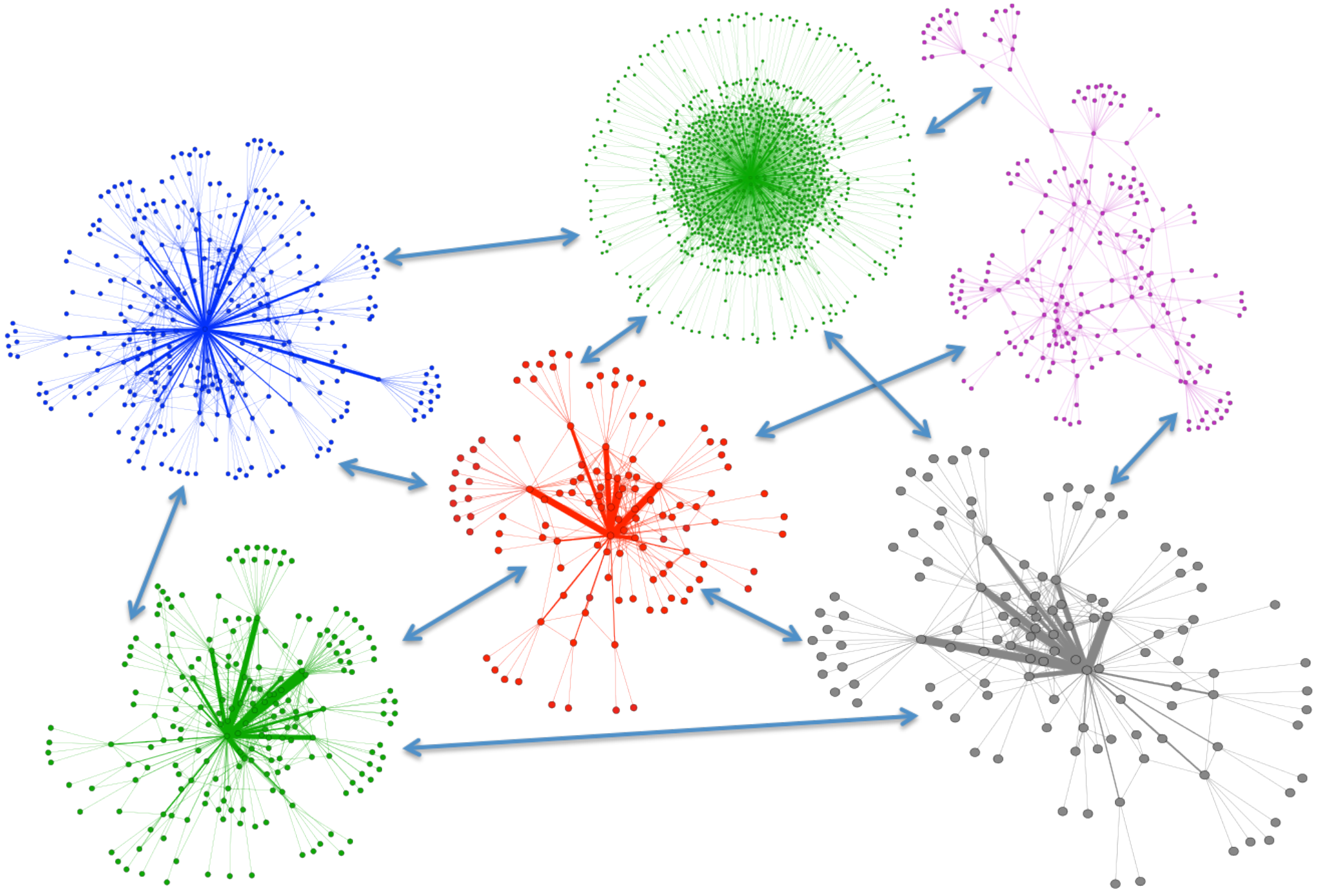}
  \caption{The Quantum Internet. Different networks are operated under different technologies. That includes the physical link architecture, data link protocol, routing protocol and any other necessary technology layered as in the OSI reference model~\cite{zimmermann1980osi}. Connecting those networks forms the Quantum Internet.}
  \label{QuantumInternet}
\end{figure}

The Quantum Internet will be a world-wide network interconnecting diverse quantum networks, both small scale and large scale~\cite{Kimble2008,VanMeter:2014:QN:2683776, Wehnereaam9288,irtf-qirg-principles-00}.
These independent, interconnected networks are utilizing different technologies and managed by different organizations (see Fig.~\ref{QuantumInternet}), known as Autonomous Systems in the classical Internet.
The role of such a network is similar to the classical Internet: to provide to users a quantum information service between arbitrary nodes.

The Quantum Internet brings us new capabilities that fundamentally cannot be reproduced by classical technologies.
Some of the well-known examples include \emph{quantum key distribution} (QKD) to securely share strings of random, secret classical bits suitable for encrypting messages~\cite{BENNETT20147, PhysRevLett.68.557, PhysRevLett.67.661, TGW},
accurately synchronizing clocks over a network~\cite{PhysRevLett.85.2010, PhysRevLett.85.2006}, distributed computing such as the secure delegate quantum computing service named \emph{quantum blind computing}~\cite{Broadbent2010, 2013arXiv1306.3664C},
and other cases involving more than one quantum computer working on difficult problems that cannot be solved by classical supercomputers.

Quantum repeaters, introduced by Briegel et al. in 1998~\cite{Briegel1998}, are the core idea of a robust quantum network.
A quantum repeater node has four main roles.
First, a repeater needs to be able to physically create,
distribute and store entangled resources between neighbors~\cite{PhysRevLett.59.2044, jones2015design}.
The second role is to manage errors on qubits.
Errors can be corrected via quantum error correction~\cite{PhysRevLett.104.180503, PhysRevA.79.032325},
or detected and discarded from the system via quantum purification~\cite{PhysRevLett.81.5932}.
Third, the node connects stored resources to increase the span of entangled states over a multi-hop route~\cite{PhysRevA.93.042338, VanMeter2013, 2017arXiv170807142P}, typically via entanglement swapping~\cite{PhysRevLett.71.4287}.
Lastly, each node needs to participate in management of the network.

Quantum systems are inherently noisy. Knill and Laflamme discussed the imperfections of quantum computer devices,
and introduced a fault-tolerant quantum communication scheme in 1996~\cite{1996quant.ph..8012K}. Later, Muralidharan et al. categorized quantum repeaters,
based on capabilities, into three generation classes~\cite{Muralidharan2016}.
The 1st generation quantum repeater network works based on Purify-and-Swap~\cite{Briegel1999, 2001Natur.414..413D, 2007PNAS..10417291J}.
The main task of a repeater node is to perform purification to detect and discard erroneous resources,
and to perform entanglement swapping to connect non-adjacent nodes.
While this scheme is relatively simple and straightforward, its capability could strictly depend on the distance,
mainly limited by classical latencies for receiving acknowledgements regarding purification and entanglement swapping.
Its performance is also known to be limited by memory lifetime~\cite{hartmann06}.
The 2nd generation utilizes encoded Bell pairs prepared between adjacent nodes,
and performs quantum error correction~\cite{PhysRevA.79.032325, PhysRevLett.104.180503, 1367-2630-15-2-023012}.
Swapping is done at the logical level for~\cite{PhysRevA.79.032325}, via an extended surface for~\cite{PhysRevLett.104.180503}.
In the 3rd generation, quantum states are directly encoded to a block of physical qubits that will be sent through the channel.
The receiver node can correct errors using the received physical qubits.
The 3rd generation is very similar to the 2nd generation, but requires a very high entanglement success rate.
These two generations are semantically identical but temporal behavior of the 3rd generation is more like classical packet forwarding networks.

\begin{figure}[htbp]
  \center
  \includegraphics[keepaspectratio,scale=0.23]{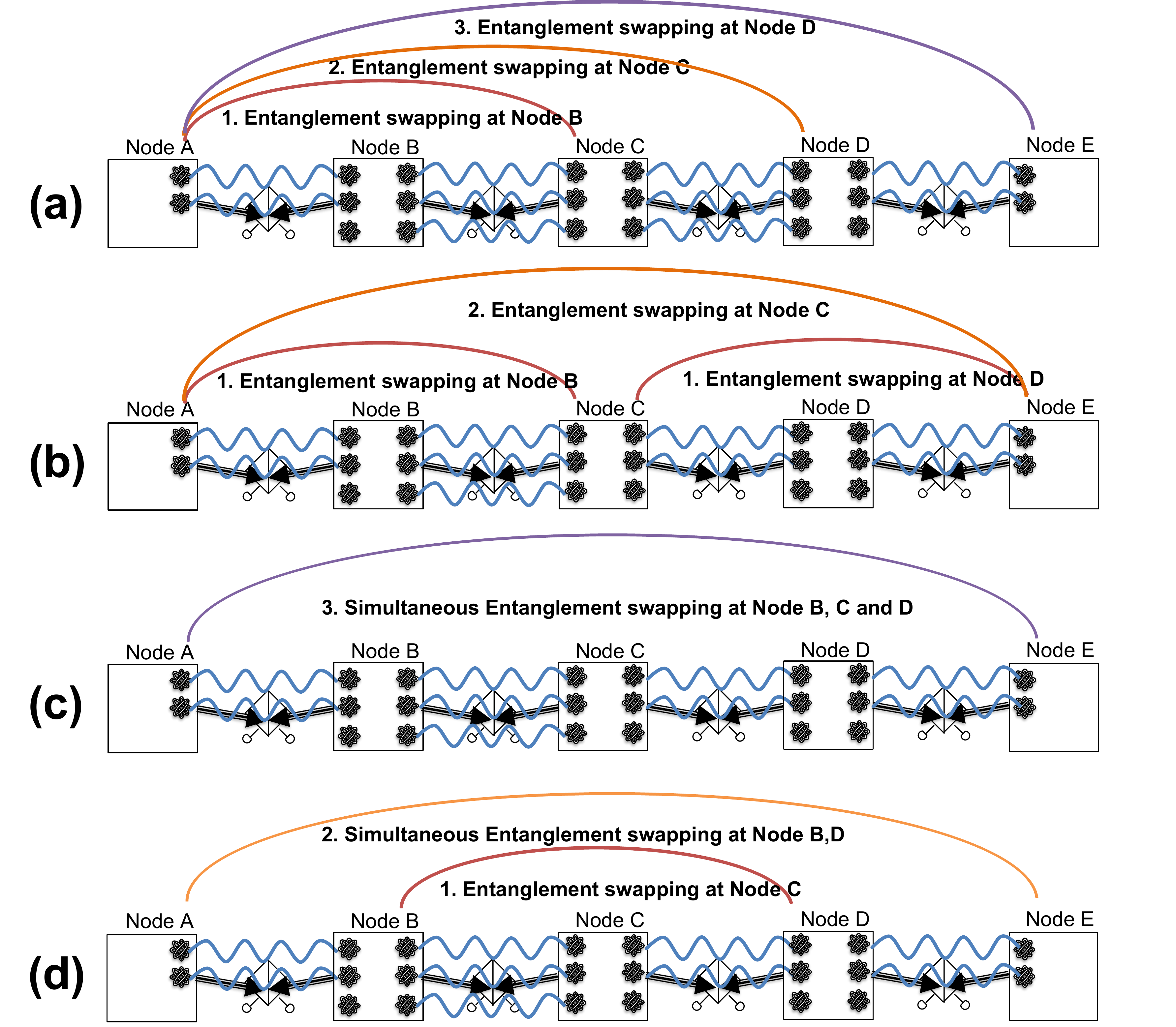}
  \caption{Examples of entanglement swapping breakdowns without multi-hop purification (each link may require purification).
  (a) Perform entanglement swapping from left to right. This keeps the information flow consistent, but requires many steps.
  (b) Node B and Node D perform entanglement swapping first, and Node C connects the end nodes afterwards. This has one less step than the previous one.
  (c) All middle nodes perform entanglement swapping simultaneously. This requires only one step, but it may work poorly with low quality links.
  (d) Ad hoc. Entanglement swapping is performed as soon as resources are available. The number of steps varies from execution to execution.}
  \label{PathBreakDown}
\end{figure}

In all three generations, establishing an end-to-end quantum connection requires knowledge regarding the links, and cannot be accomplished without coordinating quantum operations among the nodes involved~\cite{van-meter-qirg-quantum-connection-setup-00}.
The link bootstrapping protocol is capable of quantifying the achievable link fidelity~\cite{0ffddf3a0689425fb6b77dd41687a25d}.
Quantified link fidelity can be used for different purposes such as for quantum routing and end-to-end connection setup.
For example, the order of performing entanglement swapping impacts the output fidelity,
even over the same linear path, when links have different capabilities~\cite{VanMeter2013} (see Fig.~\ref{PathBreakDown}).

In this paper, we introduce the RuleSet-based communication protocol,
which we use to enforce consistency in operations over a particular connection without the need to exchange classical messages before each operation.
This allows each node to make coordinated decisions autonomously, reducing wait round-trip times.
Using RuleSets, we simulate quantum link bootstrapping that consists of
recurrent purifications~\cite{2007RPPh70.1381D} and density matrix reconstruction via link-level tomography~\cite{2005AAMOP..52..105A},
over a quantum repeater network running under specific data link protocols based on~\cite{2016NJPh...18h3015J}
with currently available hardware specifications (see Table~\ref{err} in section \ref{Simulation} for details).
We also estimate the link throughput from the bootstrapping process time relative to the measured Bell pairs.
We execute the recurrent purification based on the standard Single-selection or on Double-selection~\cite{fujii:PhysRevA.80.042308},
and find that switching from Double-selection to Single-selection in the middle of the protocol offers an advantage in terms of fidelity and throughput when resources between adjacent nodes have low fidelity.

\section{Link architectures}
Jones \emph{et al.} introduced three different quantum data link protocols, MeetInTheMiddle, SenderReceiver and MidpointSource, for distinct physical layer architectures~\cite{jones2015design}.
In this paper, we leave MidpointSource as future work, because the hardware requirements are noticeably different from to the other two.
We briefly explain the two protocols we focus on, MeetInTheMiddle and SenderReceiver, below.

\subsection{MeetInTheMiddle}
\begin{figure}[htbp]
  \center
  \includegraphics[keepaspectratio,scale=0.33]{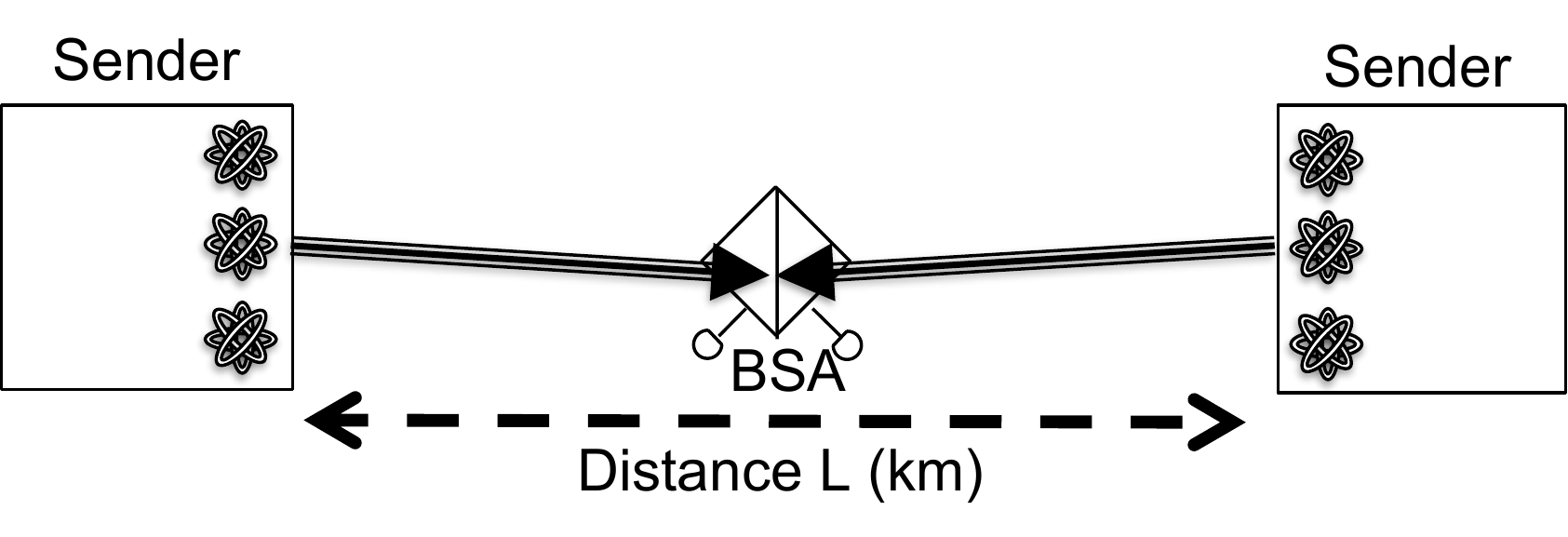}
  \caption{The MeetInTheMiddle model. In the real world, the position of the BSA could be limited for various reasons, such as geographical constraints. }
  \label{MeetInTheMiddle}
\end{figure}

A link based on MeetInTheMiddle involves two repeater nodes with one Bell state analyzer (BSA) in the middle (see Fig.~\ref{MeetInTheMiddle}).
Each node is connected to the BSA via a quantum channel and a classical channel.
The nodes are separated by a distance \(L\), with an arbitrary positioning of the BSA along the channel.
This is commonly implemented in the real world~\cite{hensen2015loophole}.
In this model, nodes need to synchronize their operations, so that the emitted photons arrive at the BSA simultaneously.
The BSA stochastically succeeds or fails in entangling the qubits,
and acknowledges the result to both nodes for each attempt.
Nodes consume all entangled qubits every round,
reinitialize all of their memory qubits and restart the cycle from the beginning.

A prototype link-layer protocol for a MeetInTheMiddle link, with few-qubits quantum processors, is discussed by Dahlberg \emph{et al.}~\cite{2019arXiv190309778D}.
The link-layer protocol is responsible for generating robust resources, and estimating the fidelity.
Generated resources will be allocated to the upper layer dedicated for networking.

\subsection{SenderReceiver}
\begin{figure}[htbp]
  \center
  \includegraphics[keepaspectratio,scale=0.33]{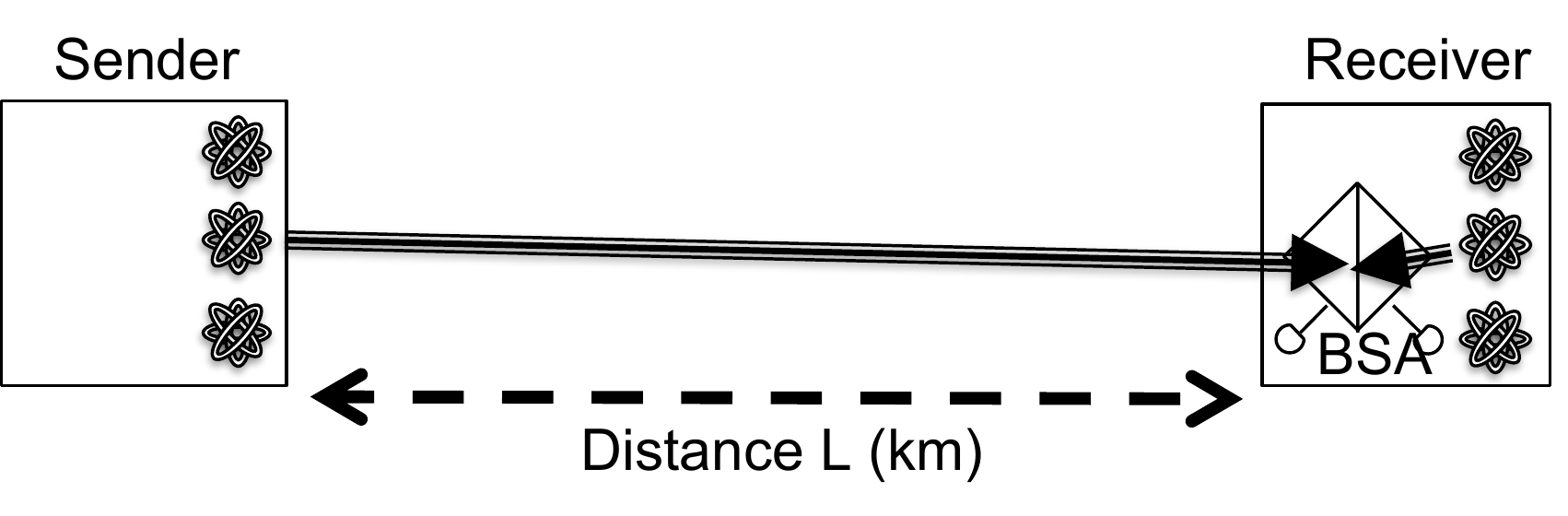}
  \caption{The SenderReceiver model. }
  \label{SenderReceiver}
\end{figure}

SenderReceiver is similar to MeetInTheMiddle, but the BSA module is installed inside one endpoint (see Fig.~\ref{SenderReceiver}).
Therefore, one node acts as a sender, while the other node with the BSA acts as a receiver.
The behavior of the nodes is similar to the ones in the MeetInTheMiddle model,
but the receiver is capable of resetting memory qubits in real time at every attempt,
because it can immediately refer to the entanglement success/failure results locally.
Therefore, the receiver can effectively utilize its memories, which will be helpful when its buffer size is smaller than that of the sender's~\cite{2010NaPho...4..792M,Aparicio2011}.
Implementing the BSA in a node directly also reduces the installation cost.
On the other hand, because the acknowledgement needs to be sent from the receiver to the sender,
compared to MeetInTheMiddle with the same distance \(L\), classical latency increases by up to a factor of two.

\section{Protocol design}
In this section, we will explain the RuleSet-based quantum networking protocol and the data link protocol, we use for the network simulation.

\subsection{RuleSet}
\begin{figure}[htbp]
  \center
  \includegraphics[keepaspectratio,scale=0.27]{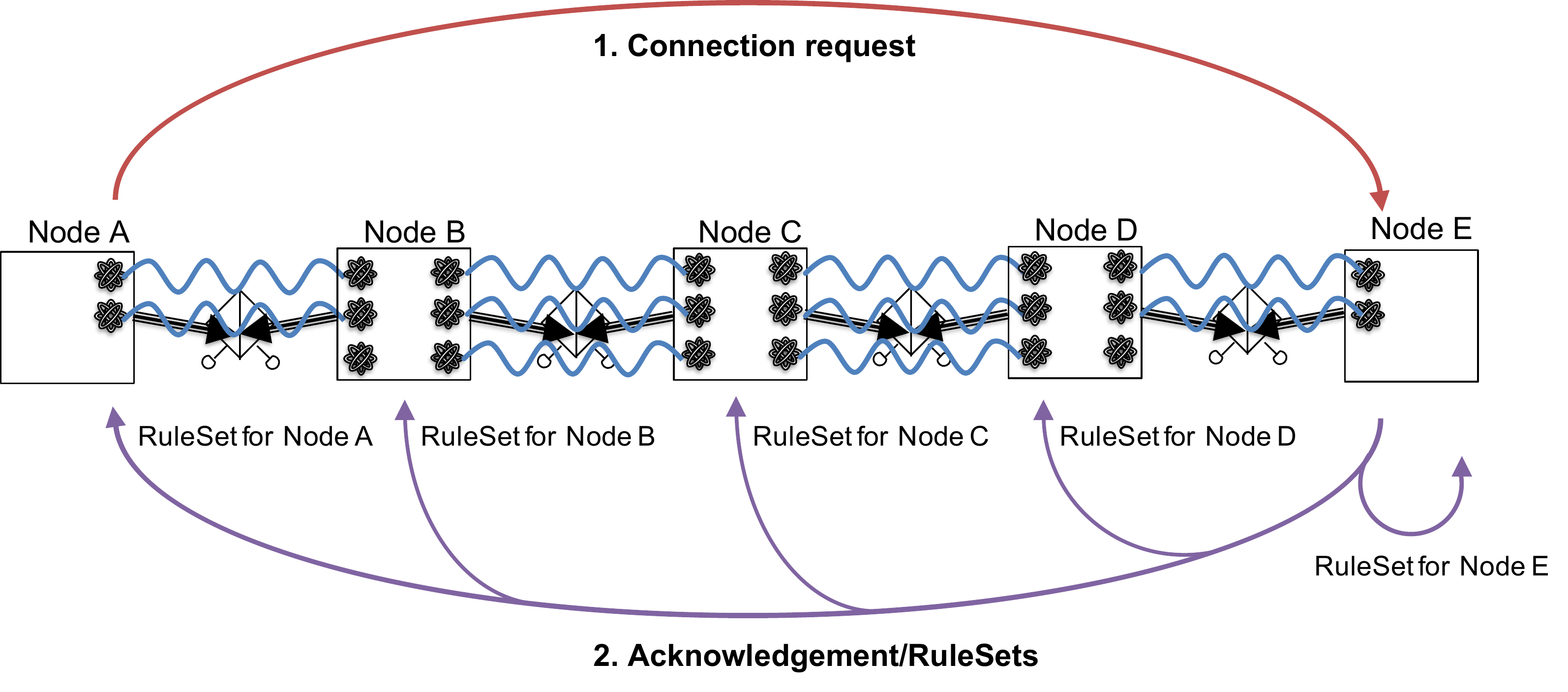}
  \caption{Node A generating and distributing RuleSets. In this example, there are five nodes in total, which requires a total of five distinct RuleSets.}
  \label{RuleSetDistribution}
\end{figure}

Coordinated operation is a prerequisite functionality for quantum networking over both the link layer and the network layer~\cite{zimmermann1980osi}.
When nodes perform a particular operation using pre-shared entangled resources,
participants are assumed to perform the correct operation targeting the appropriate resources.
Such consistency should not be achieved by exchanging messages with each other, especially over long distances due to the latency incurred.
In this section, we introduce the concept of {\it RuleSet} for supporting quantum networking,
which allows us to synchronize operations over a network with minimal classical message transmission.
If a single connection follows a route involving $n$ nodes, the source node requests the destination node to generate $n$ RuleSets,
 and distributes them to all $n$ nodes respectively (see Fig.~\ref{RuleSetDistribution})~\cite{van-meter-qirg-quantum-connection-setup-00}.

A RuleSet is an object consisting of one or more {\it Rules}, each holding a {\it Condition} and an {\it Action}.
A Condition may have one or more {\it Clauses}, each of which is a conditional statement.
An Action holds a list of operations that a node needs to perform in order to accomplish a single task,
which may be for example, entanglement swapping, purification, measurement for tomography, etc.,
and is invoked if and only if each Clause in the corresponding Condition is fulfilled.
When the task involves at least one resource, the oldest available resource will be picked from the allocated set.
An Action may also generate a message to another node, lock and reinitialize resources.
In our current implementation,
a RuleSet also has a {\it Termination Condition}, which is a simple counter for determining when to discard the RuleSet and discontinue the connection.
This may also have multiple Clauses.

\begin{figure}[htbp]
  \center
  \includegraphics[keepaspectratio,scale=0.3]{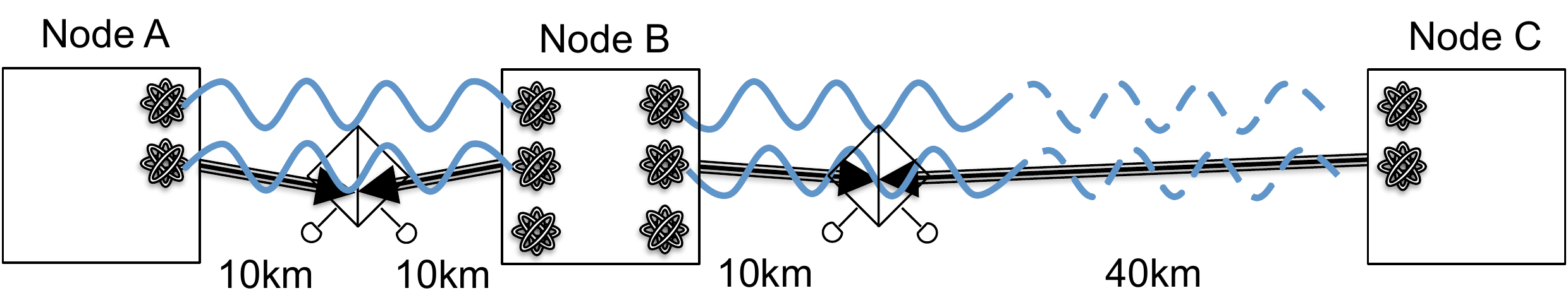}
  \caption{Bias in knowledge regarding the state of resources.  Node C receives the success/failure acknowledgement later than Node A and B.}
  \label{QI_inconsistency}
\end{figure}

RuleSets will be managed and executed by the RuleEngine,
which is a software module installed in all quantum nodes that is responsible for interpreting the RuleSet instructions, and executing them in real time.
The RuleEngine is event-driven, where the event may be a classical packet arrival, new resource allocation or RuleSet timeout.
If a single RuleSet consists of multiple Rules, the RuleEngine refers to each Rule based on top-down strategy, invoking Actions from top to bottom sequentially.
That is, if a node completes the action in the first Rule on qubit A, if any, qubit A will be reassigned for the use of the second Rule in the same RuleSet.
Therefore, newly generated resources are first assigned to the first Rule in the set.
They must complete the first rule before being reassigned to the second Rule.
Timeout may be performed by tracking how well resources are upgraded to the upper Rules.
This RuleSet level resource allocation can be directly translated into multiplexing schemes, such as buffer space multiplexing~\cite{Aparicio2011}.
This top-down approach is also a simple solution to overcome the bias in knowledge regarding the resources across a network (see  Fig.~\ref{QI_inconsistency}).

All RuleSets for the same connection share the same RuleSet identifier ({\it RuleSetID}).
The RuleSet identifier is generated via a hash function using the time when the RuleSet was generated,
the IP address of the node which generated the RuleSet, and a random number as a seed to avoid global conflicts.
Rules are also indexed ({\it RuleID}) inside the RuleSet. Each action holds a counter ({\it ActionIndex}), where the counter is incremented whenever the action has been performed.
These identifiers are used, for example, when a node needs to share the measurement result with another node for link-level tomography.
Whenever an Action of a RuleSet requires a node to transmit a classical packet to another node, such as the measurement result,
it encloses the identifiers in the packet, so that the receiving node can uniquely identify and pair its own measurement result with the received one accordingly.

\begin{figure}[htbp]
  \center
  \includegraphics[keepaspectratio,scale=0.42]{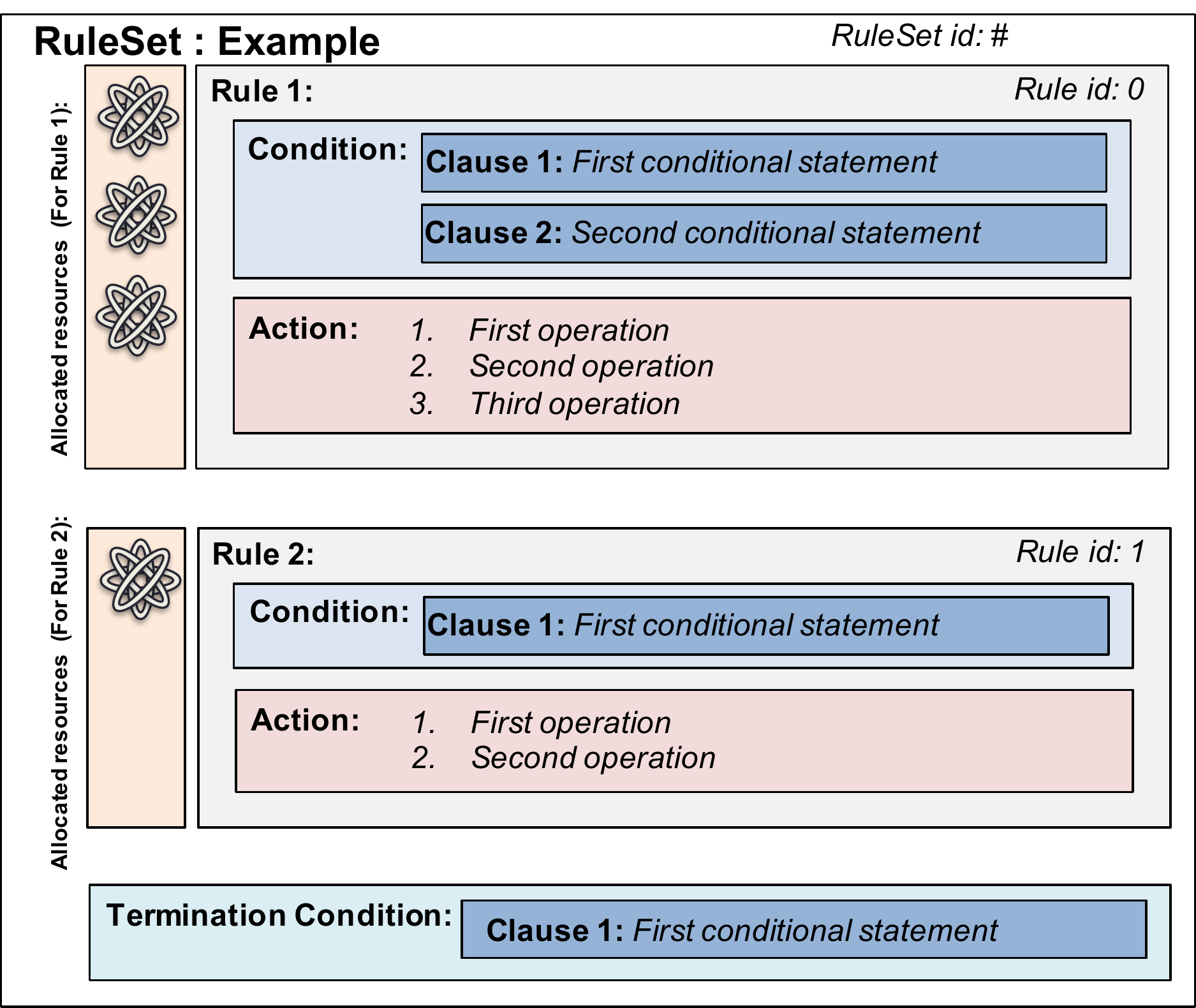}
  \caption{An example structure of a RuleSet. The resource allocated in the second Rule has been already passed through the first.}
  \label{RuleSetExample}
\end{figure}

An example structure of a RuleSet composed of two Rules is shown in Fig.~\ref{RuleSetExample}.
The pseudocode of a Rule is provided in section \ref{Simulation}.

\subsection{Data link protocol}

The core concept of the data link protocols introduced in~\cite{Codey2016} will be kept untouched.
For simplicity, in this paper, we only consider two models, MeetInTheMiddle and SenderReceiver.
MidpointSource remains as future work.
As in~\cite{Codey2016}, we do not elaborate how to synchronize clocks across a network,
but will start with the coordination of photon emissions between adjacent nodes.

\begin{figure}[htbp]
  \center
  \includegraphics[keepaspectratio,scale=0.31]{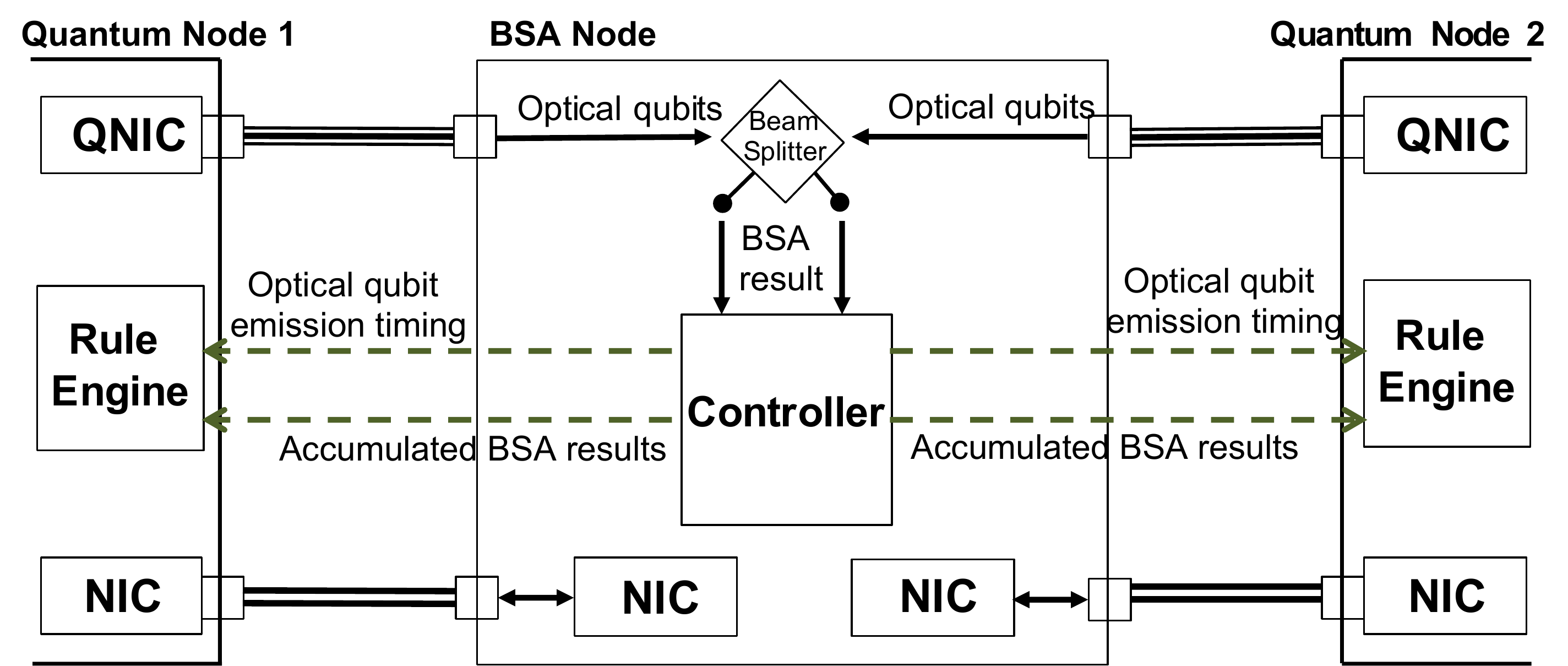}
  \caption{BSA node architecture. Classical communications, shown as dashed arrows, will be sent through the classical channel via NICs.}
  \label{BSAnode}
\end{figure}

We add some supplemental components for the network simulation.
Coordinating the arrival time of photons at the BSA via messaging between the two nodes with memory is difficult.
Instead, we make the BSA node responsible for coordinating the entanglement generation.
To start, an activated quantum node classically transmits a {\it Boot Up Notification} to all neighboring BSA nodes.
For a SenderReceiver link, the sender transmits the message towards the receiver.
Once received, the BSA node calculates the photon emission timings and the corresponding burst rate for neighboring nodes,
based on the quantum channel lengths and its own single photon detector recovery time.
Such information will be classically forwarded to the neighboring quantum nodes to start the entanglement generation.
The quantum nodes also need to classically notify the BSA node of the end of the burst.
Reception triggers the BSA to recalculate the emission timing for the next round,
and returns the information together with a list of {\it success/failure for transmission \( i\)}.
We buffer the success/failure results, and send them as a single packet to prevent overflowing the classical channel.
As in Fig.~\ref{BSAnode}, each end of a classical channel is connected to a network interface card (NIC),
and each end of a quantum channel is connected to a quantum network interface card (QNIC), which holds the memory qubits.
Optical qubits arrive at the beamsplitter from QNICs via quantum channels, and success/failure results will be accumulated by the BSA controller,
which also controls the packet transmissions.
For SenderReceiver links, the BSA module is implemented inside the receiver's QNIC.

\section{Analysis}
\label{Simulation}

In this section, we study the capability of a first generation quantum repeater link, with a limited amount of resources,
 and under noisy conditions generated by various hardware imperfections.
Simulated error sources include quantum memories, quantum channels, quantum gates and single photon detector darkcounts.
Our Markov-Chain Monte-Carlo simulation uses similar parameters as in~\cite{2018arXiv180900364R} but with larger memory buffers (see Table~\ref{err} for details).
We use the memoryless Markov-Chain in the simulation to dynamically model errors on memory qubits independently, based on a given lifetime $T_1$~\cite{van-meter19:errm} (for details, see Appendix~\ref{memoryerror}).
Other error sources, such as the gate error, are simulated via a random selection with static probabilities.
In order to reduce the computation time, we only propagate Pauli errors through circuits as in~\cite{PhysRevA.86.032331, PhysRevA.97.062328}.
The simulated purification circuit, therefore, is incapable of stochastically detecting excited/relaxed/completely mixed errors.
Hence, our simulation generates a pessimistic output fidelity, and an optimistic output resource generation rate.
To concretely identify the real impact of imperfect quantum systems, we assume ideal classical communication channels.
Because classical communication latencies cover the majority of the simulated time, we assume negligible gate times.
Furthermore, multi-qubit operations can also be performed between arbitrary qubits across QNICs within the same node.
For each data point in the figures, a total of 25 simulations has been performed to quantify the average behavior.

\begin{table}[htbp]
  \caption{Default parameters used for all simulations (unless explicitly mentioned).}
  \label{err}
  \begin{ruledtabular}
    \begin{tabular}{lr}
    \multicolumn{1}{c}{\bf Key} & \multicolumn{1}{r}{\bf Value}\\
    \hline
     Fiber refractive index & $1.44$~\cite{fiber} \\
    Fiber Pauli error rate (per km) & $0.03$ \footnote{\label{pauli}Including X, Y and Z error.}\\
    Fiber photon loss rate (per km) & 0.04501 \footnote{\label{db}Equivalent to 0.2dB/km.}~\cite{fiber}\\
    Memory Pauli error rate (per sec) & 1/3 \footref{pauli}~\cite{Maurer1283}\\
    Memory lifetime & 50ms \footnote{\label{memory}Memory lifetime can be, for example, up to the order to seconds~\cite{Maurer1283} depending on the system. We use an artificial value of 50ms for the simulation. The ratio of excitation and relaxation probability is set to 100:1.}\\
    Emission probability into the zero phonon line & 0.46 \footnote{\label{emission}Photon emission probability (memory-to-fiber) is the product of the emission probability into the zero phonon line and the collection efficiency.}~\cite{PhysRevX.7.031040} \\
    Photon collection efficiency & 0.49 \footref{emission}~\cite{HensenNature, doi:10.1063.1.5001144}\\
    Photon detector efficiency & 0.8~\cite{doi:10.1063.1.5001144} \\
    Photon detector darkcount rate (per sec) & 10~\cite{doi:10.1063.1.5001144} \\
    Photon detector recovery time  & 1ns~\cite{SNSPDs} \\
    Single-qubit gate error rate  & 0.0005 \footref{pauli}~\cite{PhysRevX.6.021040} \\
    Multi-qubit gate error rate & 0.02 \footref{pauli}~\cite{PCNature} \\
    Measurement error rate & 0.05 \footref{pauli}\cite{Kalb928} \\
    Number of memory qubits (per QNIC) & 100
    \end{tabular}
  \end{ruledtabular}
\end{table}

\subsection{Performing simple tomography with different numbers of measurements}
Performing quantum state tomography with more measurements typically results in a greater accuracy of the reconstructed density matrix.
Although the required accuracy may depend on the demanded precision of a particular application,
it is unlikely to perform an extremely large set of measurements solely for the reconstruction for purposes such as real time quantum channel monitoring.
We first investigate how the number of measurements for the tomography, $N_{M}$,
 impacts the accuracy of the \emph{reconstructed fidelity}, $F_{r}$, between adjacent nodes.
Here, $\mathrm{F_{r}} = \mathrm{Tr}[\rho_{r}\rho_{i}]$,
where $\rho_{i}$ is the ideal density matrix of a Bell pair $\ket{\Phi^{+}}$
and $\rho_{r}$ is the density matrix reconstructed through the RuleSet-based quantum link-level tomography.
This particular RuleSet consists of a single Rule composed of two Clauses and an Action.
The first Clause, {\it MeasurementConditionClause}, tracks how many measurements have been performed.
The second Clause, {\it ResourceConditionClause}, checks the availability of resources (see Algorithm \ref{ResourceClause}).
The Action performs the measurement for the link-level tomography (see Algorithm \ref{TomographyAction}).
In this case, the Termination Condition also tracks the number of performed measurements, and stops the execution when the requirement is fulfilled.

\begin{algorithm}[H]
  \algrenewcommand\algorithmicrequire{\textbf{Clause:}}
  \caption{ResourceConditionClause}
  \label{ResourceClause}
  \begin{algorithmic}[1]
    \Statex {\bf{This Clause checks if enough resources are available for the corresponding Action.}}
    \Statex {Input: resourceList $\gets$ List of allocated resources for the Rule. A resource may be locked when it is waiting for the classical packet to arrive, such as for purification.}
    \Statex {Output: enoughResources $\gets$ A boolean value.}
    \Statex {}
    \Procedure{ResourceConditionClause}{resourceList}
    \State {numRequired $\gets$ Number of required resources for the Action}
    \State {numFree $= 0$}
    \State {enoughResources $=$ false}
    \For {each resource in resourceList}
      \If {resource is not locked}
        \State {numFree$++$}
      \EndIf
      \If {numFree $>=$ numRequired}
        \State {enoughResources $=$ true}
        \State {break}
      \EndIf
    \EndFor
    \State {return enoughResources}
  \EndProcedure
  \end{algorithmic}
\end{algorithm}

\begin{algorithm}[H]
  \caption{TomographyAction}
  \label{TomographyAction}
  \begin{algorithmic}[1]
  \Statex {\bf{This Action performs measurement on an entangled qubit.
  The measurement result and the selected basis is stored, and sent to the partner node as a message.}}
  \Statex{Input: resourceList $\gets$ List of allocated resources for the Rule.}
  \Statex {Output: msg $\gets$ A message for another node.}
  \Statex {}
  \Require {ResourceConditionClause == true \&\& MeasurementConditionClause == true}
  \Procedure{TomographyAction}{resourceList}
    \State {self\_addr $\gets$ Node address of this RuleSet owner.}
    \State {partner\_addr $\gets$ Node address of tomography partner.}
    \State {resource = Select a free resource from resourceList}
    \State {basis = RandomBasisSelect(\{X,Y,Z\})}
    \State {outcome = resource.Measure(basis)}
    \State {removeResourceFromList(resourceList,resource)}
    \State {Data.result = \{outcome, measurementBasis\}}
    \State {Data.ruleSetId = this.RuleSetID}
    \State {Data.ruleId = this.RuleID}
    \State {Data.actionIndex = this.ActionIndex}
    \State {save(Data)}
    \State {msg.destination = partner\_addr}
    \State {msg.source = self\_addr}
    \State {msg.data = Data}
    \State {this.ActionIndex$++$}
    \State {return msg}
  \EndProcedure
  \end{algorithmic}
\end{algorithm}

\begin{figure}[htbp]
  \center
  \includegraphics[keepaspectratio,scale=0.25]{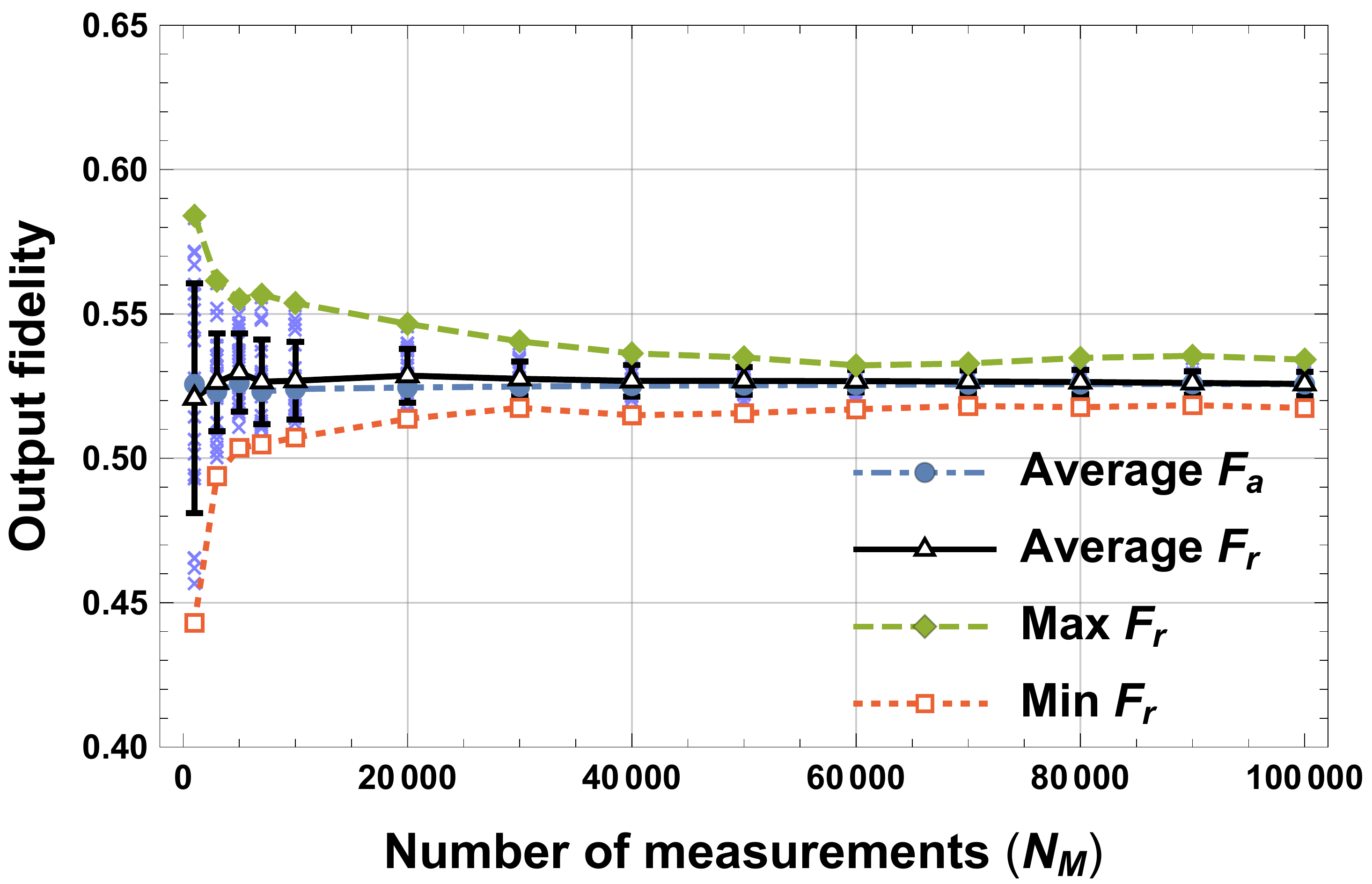}
  \caption{Impact of the number of measurements on the reconstructed fidelity error over the MeetInTheMiddle link. $F_{r}$ is the reconstructed fidelity and $F_{a}$ is the actual fidelity.}
  \label{MeasurementCount}
\end{figure}

The simulated result of a repeater network composed of two nodes based on a MeetInTheMiddle link,
each 10km from the BSA node ($L=20km$), is shown in Fig.~\ref{MeasurementCount}.
Error bars represent one standard deviation uncertainty ($\sigma$) from the average $F_{r}$.
$\mbox{Max } F_{r}$ and $\mbox{Min } F_{r}$ are the highest and lowest fidelity observed within the 25 trials.
Purple data points marked with cross between the minimum and the maximum are individual simulation results.

As in Fig.~\ref{MeasurementCount}, the reconstructed fidelity has poor accuracy with only 1,000 measurements, and slightly underestimates the value simply due to the lack of samples - $\sigma \approx 0.040$.
In order to achieve $\sigma < 0.015$, at least 7,000 measurement shots are required.
A total of 20,000 measurements is capable of reconstructing the fidelity with a standard deviation less than $1\%$,
and roughly converges to $\sigma \approx 0.004$ when $N_{M}>60,000$.

\begin{figure}[htbp]
  \center
  \includegraphics[keepaspectratio,scale=0.25]{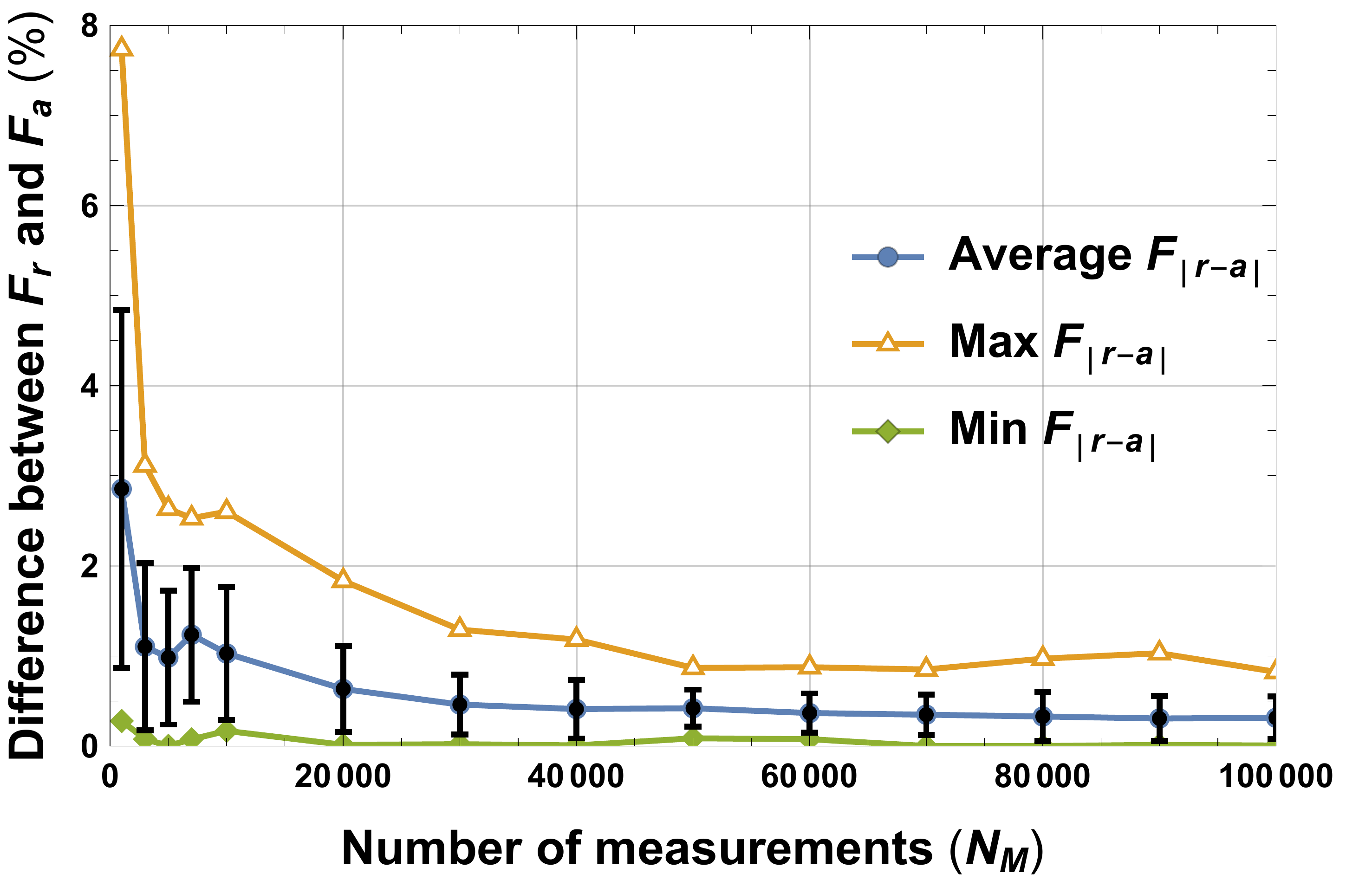}
  \caption{Change in the accuracy of the reconstructed fidelity relative to the actual fidelity.}
  \label{DifferenceF}
\end{figure}

$\mathrm{F_{a}}$ is the actual fidelity, accessible to us because this is a simulation, but of course hidden in the real world.
The change in the accuracy of the reconstructed fidelity compared to the actual fidelity is shown in Fig.~\ref{DifferenceF}.
The average difference between the reconstructed fidelity and the actual fidelity for a given $N_{M}$ is

\begin{eqnarray}
  \label{DiffF}
  \overline{F}_{\mid r-a \mid} = \frac{1}{N} \sum_{i=1}^{N} \mid F_{r(i)} - F_{a(i)} \mid,
\end{eqnarray}

where $N = 25$ is the total number of simulation trials.

The fidelity reconstruction process using only 1,000 measurement outcomes will result in about $3\%$ error from the actual value on average.
The accuracy improvement converges to $\Delta F_{\mid r-a \mid} > 0.3\%$ when $N_{M} \geq 40,000$,
in which case not much benefit can be gained from solely increasing the number of measurements.

The MeetInTheMiddle link and the SenderReceiver link have similar behavior regarding the fidelity,
with sufficiently long memory lifetime relative to the idle time caused by classical latencies.
Nevertheless, even over a single hop, the throughput strongly depends on the physical link architecture and the adopted data link protocol.
Accumulating more measurements will increase the tomography time linearly, for both links,
as the operation here consists only of measuring entangled qubits sequentially.
The entanglement generation rate over the MeetInTheMiddle link is twice of what the SenderReceiver link achieves,
simply because the BSA in the MeetInTheMiddle sits half-way through the link.
With the given parameter settings, architecture and protocols,
the MeetInTheMiddle link's Bell pair generation rate is around 6741/s (3368/s for the SenderReceiver link),
 when qubits are immediately measured when the system recognizes that they are entangled.

\subsection{Performing tomography with single shot purification over different distances}
\label{purifications}
Lengthening the channel between nodes increases the chance for optical qubits and memory qubits to experience errors.
One way to tolerate such errors is to perform quantum purification.

\begin{figure}[!htbp]
  \center
  \includegraphics[keepaspectratio,scale=0.38]{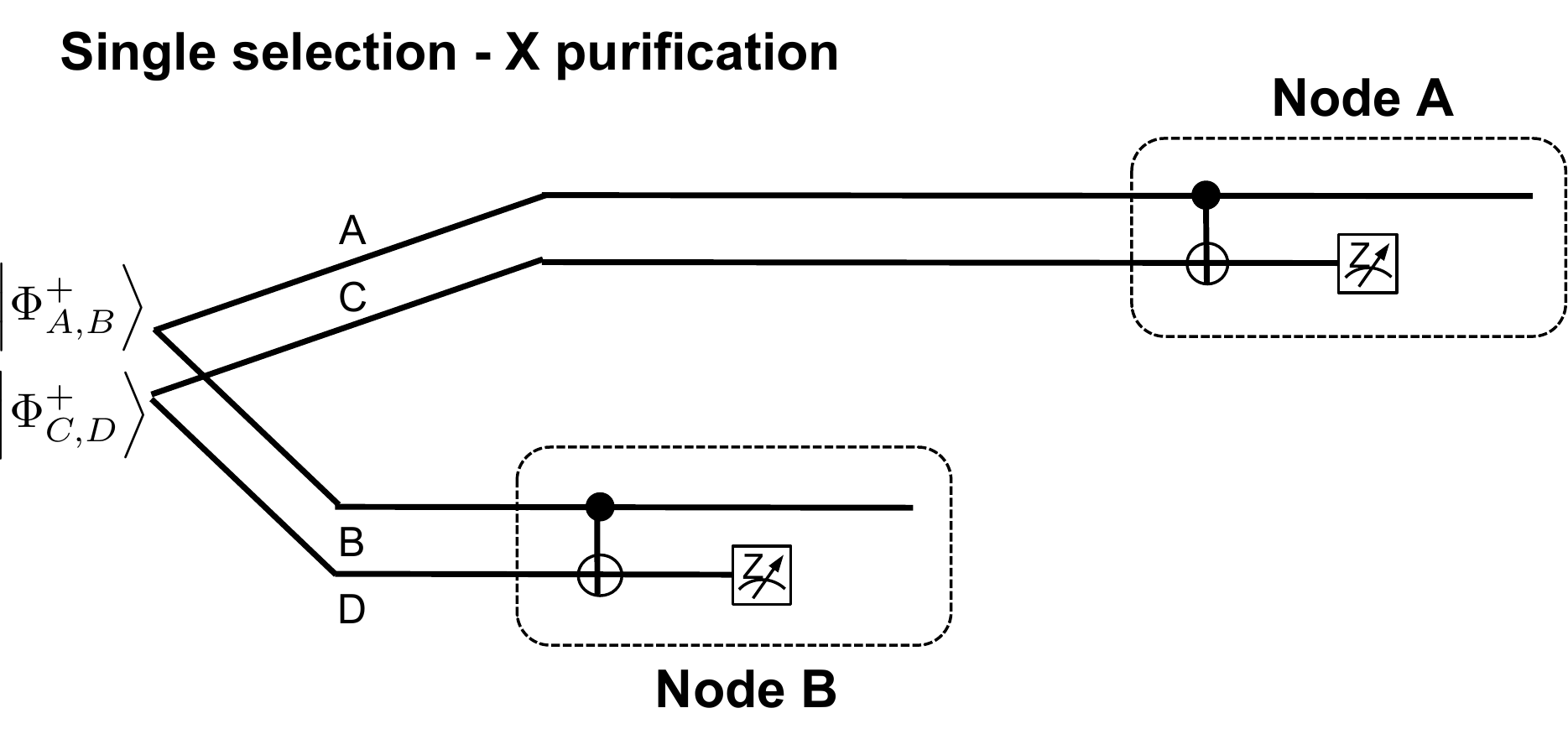}
  \caption{Single selection - Single error (X) purification (Ss-Sp). Consumes a single Bell pair $ \ket{\Psi^+_{C,D}}$ to detect the X error on $ \ket{\Psi^+_{A,B}}$.}
  \label{SS-Xp}
\end{figure}

The simplest purification scheme is the Single selection -  Single error purification (Ss-Sp)~\cite{PhysRevLett.81.5932, fujii:PhysRevA.80.042308}.
As in Fig.~\ref{SS-Xp}, the Ss-Sp requires two Bell pairs along the channel to be purified -- one Bell pair is consumed to detect the presence of an X error (or Z error) on the other Bell pair.
The resource is only purified successfully when the measurement outcomes of the consumed qubits (qubit C and qubit D in Fig.~\ref{SS-Xp}) coincide.
Otherwise, both nodes discard their resources.

\begin{figure}[!htbp]
  \center
  \includegraphics[keepaspectratio,scale=0.38]{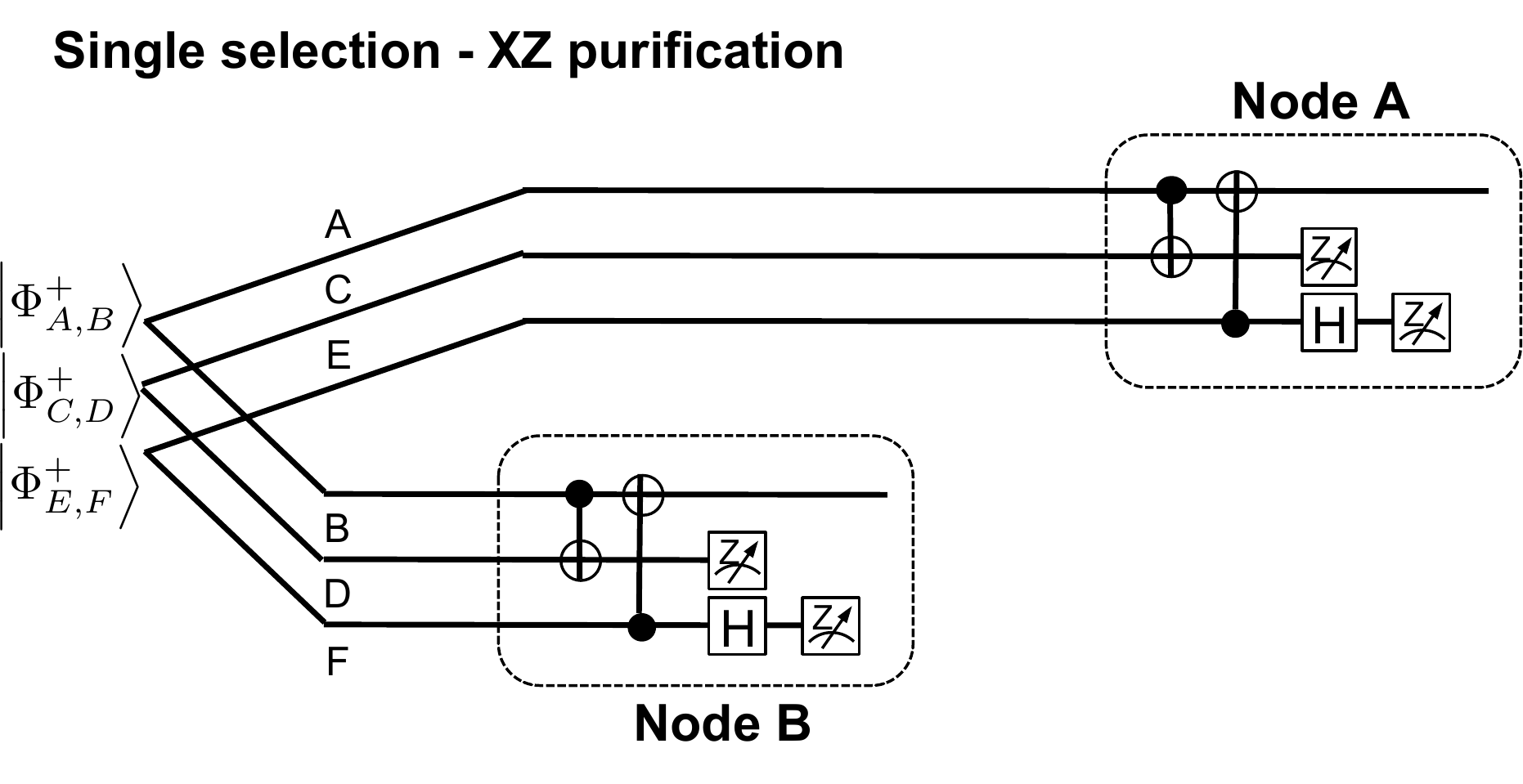}
  \caption{Single selection - Double error (XZ) purification (Ss-Dp). Consumes two Bell pairs, $ \ket{\Psi^+_{C,D}}$ and $\ket{\Psi^+_{E,F}}$ to detect X error and Z error accordingly.}
  \label{SS-XZp}
\end{figure}

By extending Ss-Sp, we can also try to purify both errors (see Fig.~\ref{SS-XZp}).
The Single selection - Double error purification (Ss-Dp) is similar to Ss-Sp,
but requires another Bell pair to detect the Z error on the resource.
However, because we perform the Z error purification after the X,
the X error can propagate to the purified resource from the consumed Bell pair.
In Fig.~\ref{SS-Xp}, the X error from qubit E and qubit F will propagate to qubit A and qubit B through the CNOT gate accordingly.

\begin{figure}[!htbp]
  \center
  \includegraphics[keepaspectratio,scale=0.38]{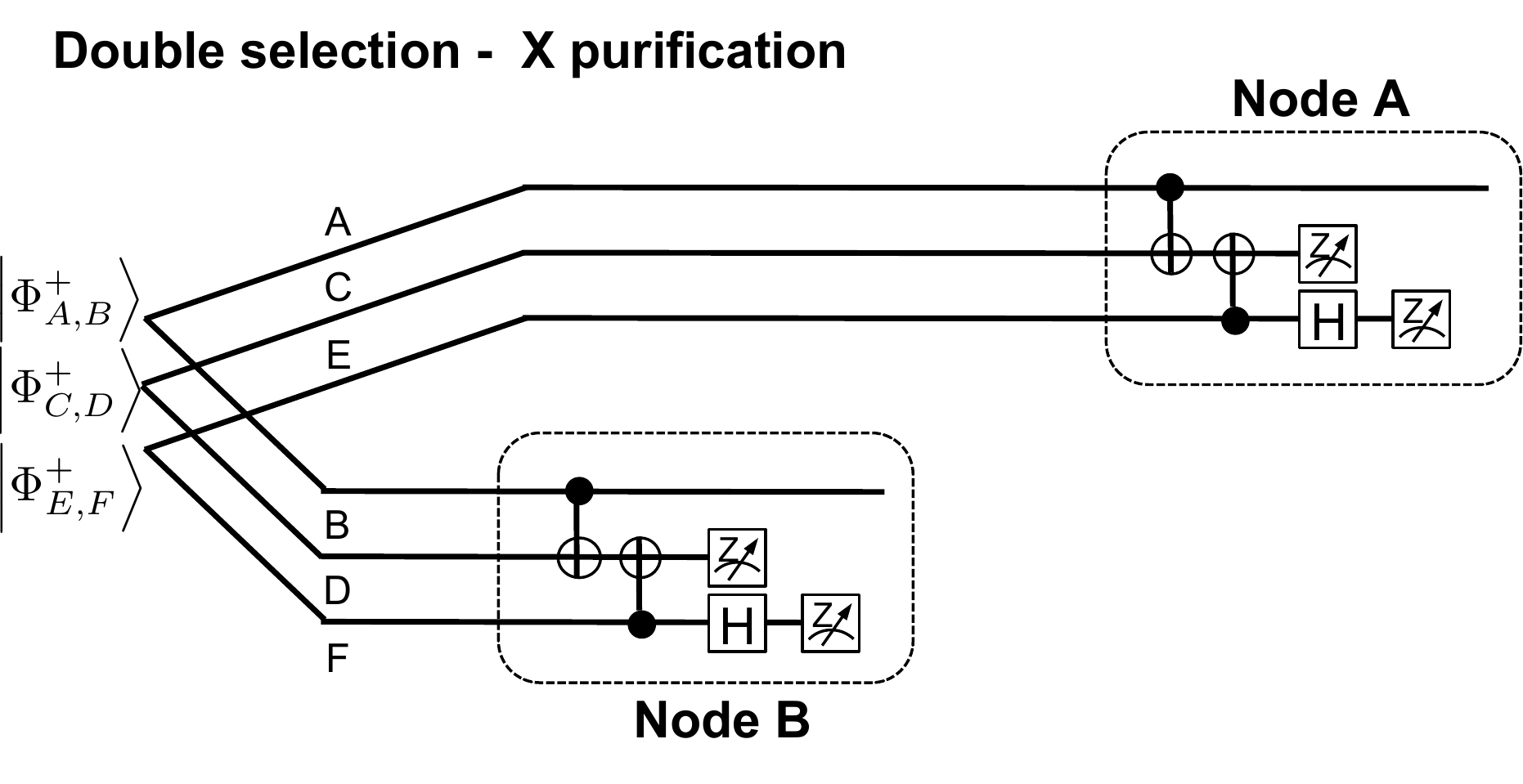}
  \caption{Double selection - Single error (X) purification (Ds-Sp).
  Consumes a single Bell pair $ \ket{\Psi^+_{C,D}}$ to detect X error on $ \ket{\Psi^+_{A,B}}$, and another Bell pair $ \ket{\Psi^+_{E,F}}$ to avoid the Z error propagation from $ \ket{\Psi^+_{C,D}}$ to $ \ket{\Psi^+_{A,B}}$.
  }
  \label{DS-Xp}
\end{figure}

The double selection method introduced by Fujii and Yamamoto in 2009~\cite{fujii:PhysRevA.80.042308} is a purification method utilizing an additional resource to improve the accuracy of post-selection by applying double verification.
The Double selection - Single error purification (Ds-Sp) does the same job as Ss-Sp,
but with an additional resource to detect the presence of a Z error on the consumed resource used for the X error purification (see Fig.~\ref{DS-Xp}).
This minimizes the error propagation to the purified resource.
Notice that any Z error that is originally present on the purified resource will not be detected.

\begin{figure}[!htbp]
  \center
  \includegraphics[keepaspectratio,scale=0.38]{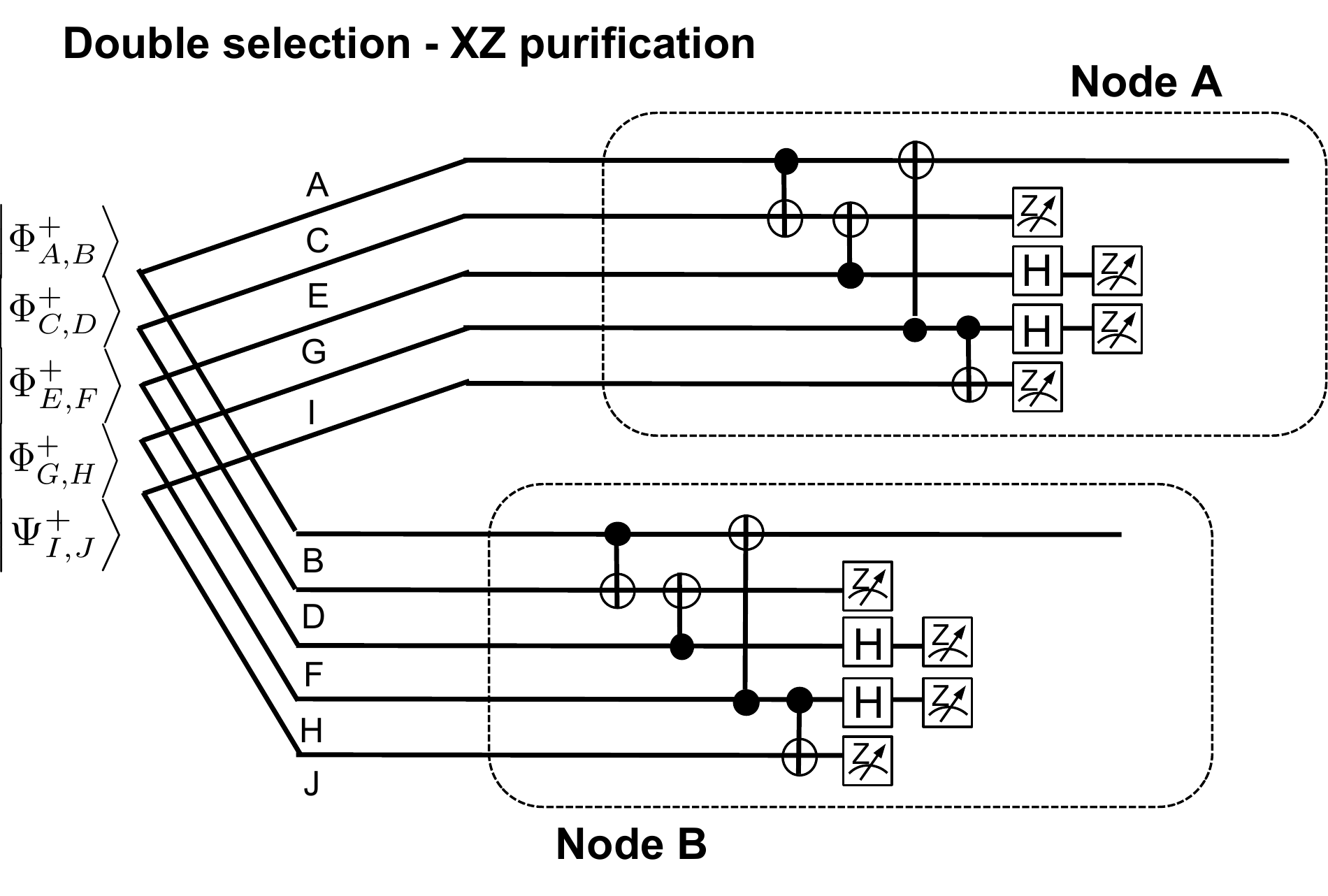}
  \caption{Double selection - Double error (XZ) purification (Ds-Dp).
  Consumes two Bell pairs, $ \ket{\Psi^+_{C,D}}$ and $ \ket{\Psi^+_{G,H}}$, to detect X error and Z error on $ \ket{\Psi^+_{A,B}}$. Each consumed Bell pair will also be verified through purification using $ \ket{\Psi^+_{E,F}}$ and $ \ket{\Psi^+_{I,J}}$ accordingly.}
  \label{DS-XZp}
\end{figure}

The double selection scheme can also be applied to the double error purification circuit (Ds-Dp).
The concept of this circuit is similar to Ds-Sp,
except we perform the double verification process on both X and Z purification to the resource we keep.
Thus, we can also minimize the error propagations that happen during the purification.
For the circuit diagram, see Fig.~\ref{DS-XZp}.

\begin{figure}[!htbp]
  \center
  \includegraphics[keepaspectratio,scale=0.25]{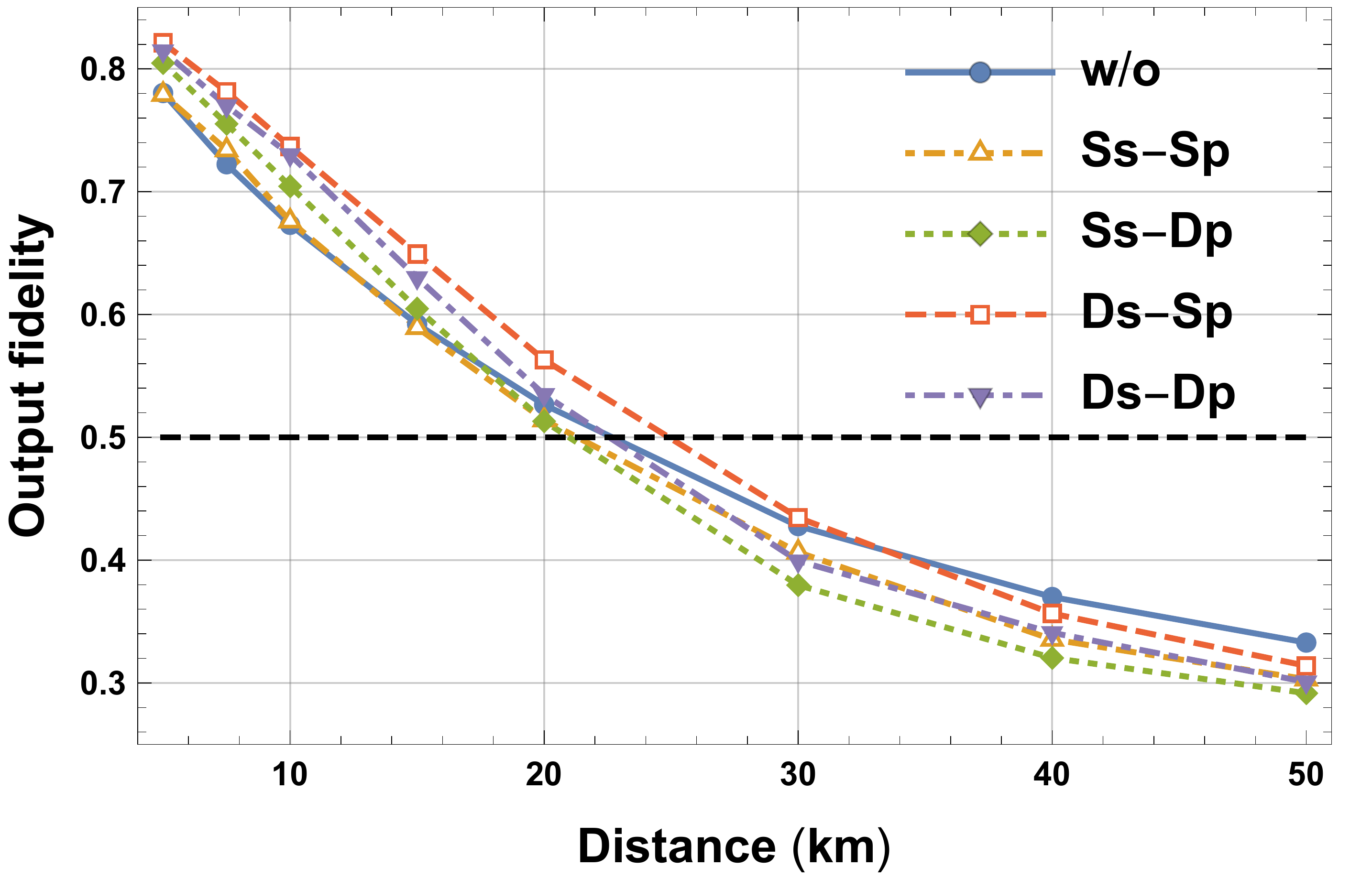}
  \caption{Impact of the MeetInTheMiddle channel length on the reconstructed fidelity with 7000 measurement outcomes w/o, with purification.
  The steep drop in the fidelity is due to the high error rate of 1\% each X, Y and Z errors per km in the quantum channel chosen for this simulation.
  }
  \label{FidelityDistance}
\end{figure}

\begin{figure}[!htbp]
  \center
  \includegraphics[keepaspectratio,scale=0.24]{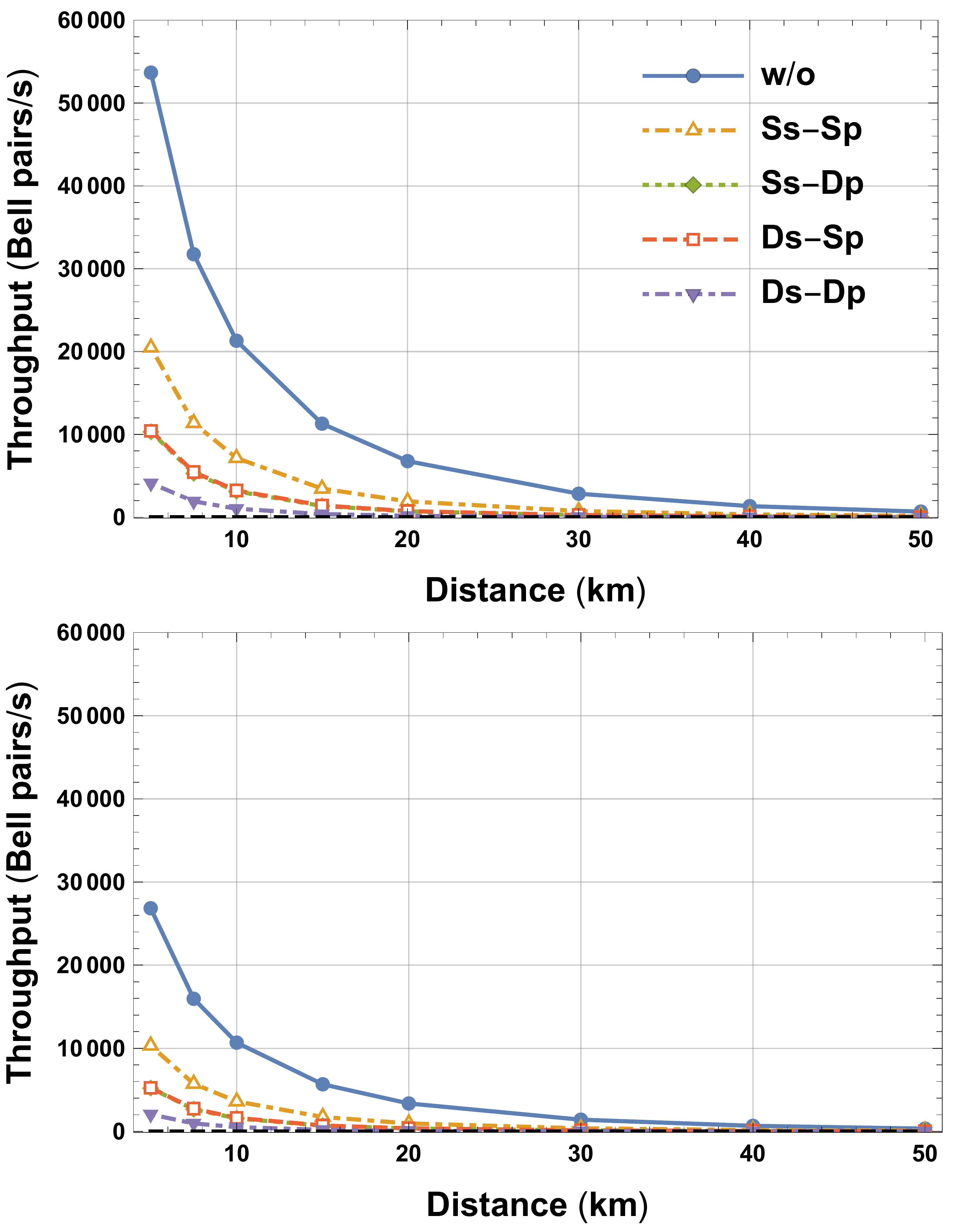}
  \caption{
  Impact of the channel distance on the channel throughput with and without performing purification.
  The throughput has been calculated as $N_{M}/T$, where $T$ is the tomography process time to complete $N_{M}$ measurements.
  (a) Throughput over the MeetInTheMiddle link.
  (b)  Throughput over the SenderReceiver link.
  }
  \label{ThroughputDistance}
\end{figure}

In this subsection, we describe simulated link-level tomography with 7000 measurements,
with and without Rules that perform one of the purification methods above beforehand, over different distances.
The simulated results of the fidelity reconstruction is shown in Fig.~\ref{FidelityDistance}, and Fig.~\ref{ThroughputDistance}.
The fidelity for SenderReceiver link is slightly worse than that of the MeetInTheMiddle link,
but the overall behavior is approximately the same.

As shown by the results in Fig.~\ref{FidelityDistance},
performing Ss-Sp once slightly improves the fidelity compared to the case without, only over relatively shorter distances,
but with a significant penalty on the throughput for both link architectures (see Fig.~\ref{ThroughputDistance}).
With a total distance $L=10km$, Ss-Sp suppresses about 5.6\% of the X errors and 5.4\% of Y errors,
but the Z error rate increases approximately 8.8\% due to the error propagation through the purification circuit.
Overall, the fidelity improves roughly by 3\%.
While the fidelity did not change dramatically,
the error distribution developed significantly different after performing the operation,
which we can still take advantage of in the recurrence purification protocol (see Section \ref{RpSec}).
By implementing double selection to the same circuit (Ds-Sp), the output fidelity improved significantly
-- about 6.4\% increase from the case without performing any purification.
The second verification decreased the Z error rate by about 4.6\% compared to the case of Ss-Sp with $L=10km$.
The Ss-Dp protocol purifies the X error first, and then the Z error afterwards.
In this case, over $L=10km$ distance,
the Z error rate decreases by 3.3\% and the Y error rate by 5.7\% -- the X error rate increases by 4.3\% due to propagation.
Simply applying double selection to this (Ds-Dp) improved the fidelity by approximately 2.6\% from Ss-Dp.
Both purifications based on double selection are capable of purifying resources under longer distances, which is roughly 20km.

With the given hardware qualities in Table~\ref{err},
Ds-Sp obtains the highest output fidelity.
The output fidelity for Ds-Dp is lower than that of Ds-Sp,
mainly because of larger $KQ$ with noisy gates, and the increase waiting time for resources, especially over long distances,
 where $KQ$ is the product of the circuit depth ($K$) and the number of qubits ($Q$)~\cite{2003PhRvA68d2322S}.
Under this scenario, because Ds-Sp only requires three resources at once, the throughput is also higher than that of Ds-Dp.

\subsection{Bootstrapping using Ruleset-based recurrence purification and tomography}
\label{RpSec}
Network operations, such as routing, often require knowledge regarding the links - e.g. fidelity, throughput, etc.
Therefore, the channel characteristics to be shared need to be acquired beforehand via quantum link bootstrapping and characterization.
In quantum networking, we commonly focus on optimizing the link fidelity for realizing a robust connection,
where the goal may be achieved by performing nested purifications.
The recurrence protocol~\cite{2007RPPh70.1381D},
performs quantum purification on top of pre-purified resources to effectively improve the fidelity
-- the required number of resources for the operation, on the other hand, increases exponentially in the number of purification rounds $N_p$,
which alone linearly affects the total idle time due to classical communication latencies.

\begin{figure}[!htbp]
  \center
  \includegraphics[keepaspectratio,scale=0.38]{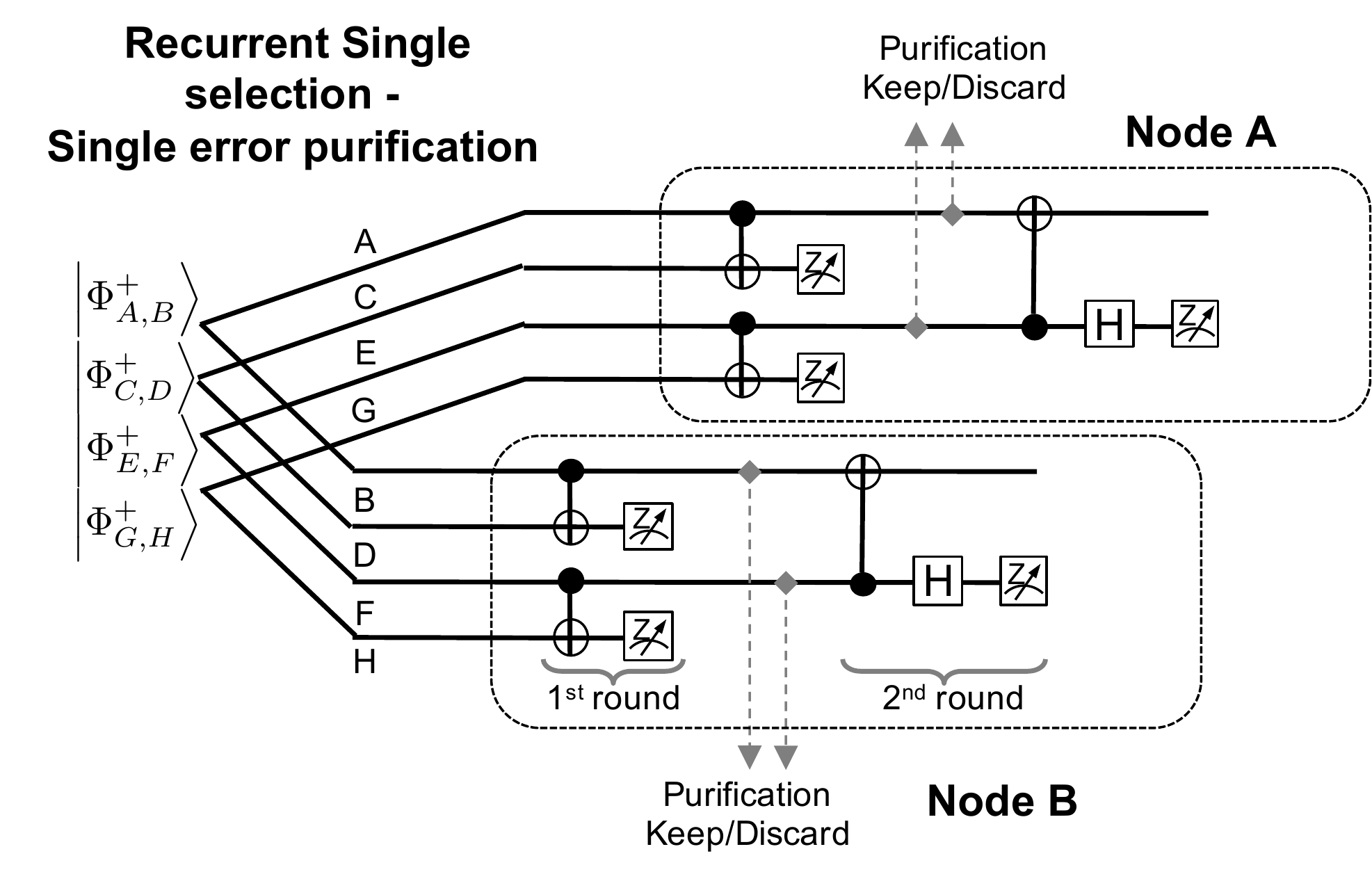}
  \caption{Recurrence purification protocol based on single selection X purification and Z purification.
  Each round of purification is a Rule in a Ruleset.
  }
  \label{Rp}
\end{figure}

In this subsection, we describe our simulation of quantum link bootstrapping to quantify the achievable link fidelity and its corresponding throughput,
using RuleSet-supported tomography with recurrence purification based on one of the four circuits shown in subsection \ref{purifications};
recurrent single selection single error purification (RSs-Sp),
recurrent single selection double error purification (RSs-Dp),
recurrent double selection single error purification (RDs-Sp) and recurrent double selection double error purification (RDs-Dp).
Each round of purification is a separate Rule in the same RuleSet.
In the given example of a two round RSs-Sp circuit shown in Fig.~\ref{Rp},
the first Rule performs X purifications to resources.
In the second Rule, we use those purified resources to perform another Z error purification.
Each round, we alternate the X purification and the Z purification to suppress the error propagation.
For double error purifications, we alternate the XZ purification and the ZX purification.
With probabilistic generation of Bell pairs and scarce memory qubits,
operations can stall in the middle, in which case the Rule waits for new resources to arrive.
Using RuleSets, the whole task is completed autonomously over a network.
The simulation continues reconstructing the density matrix every round,
and terminates the process when the same operation results in a lower reconstructed fidelity, or due to timeout.
For practical reasons, each RuleSet timeout is set to two minutes.

We assume perfect photon emissions from memories to fiber,
where in the real world, similar effects may be observed by increasing the number of memory qubits.
CNOT gates are also assumed to be ideal to solidify the real merit of recurrence protocols.
We first focus on a repeater network with a relatively short channel, where the total length is set to 10km.
Secondly, we perform similar simulations over a longer distance, $L=20km$.
For the MeetInTheMiddle link, each node is the same distance away from the BSA node.

\subsubsection{Closely spaced repeater nodes}

\begin{figure*}
   \center
   \includegraphics[keepaspectratio,scale=0.17]{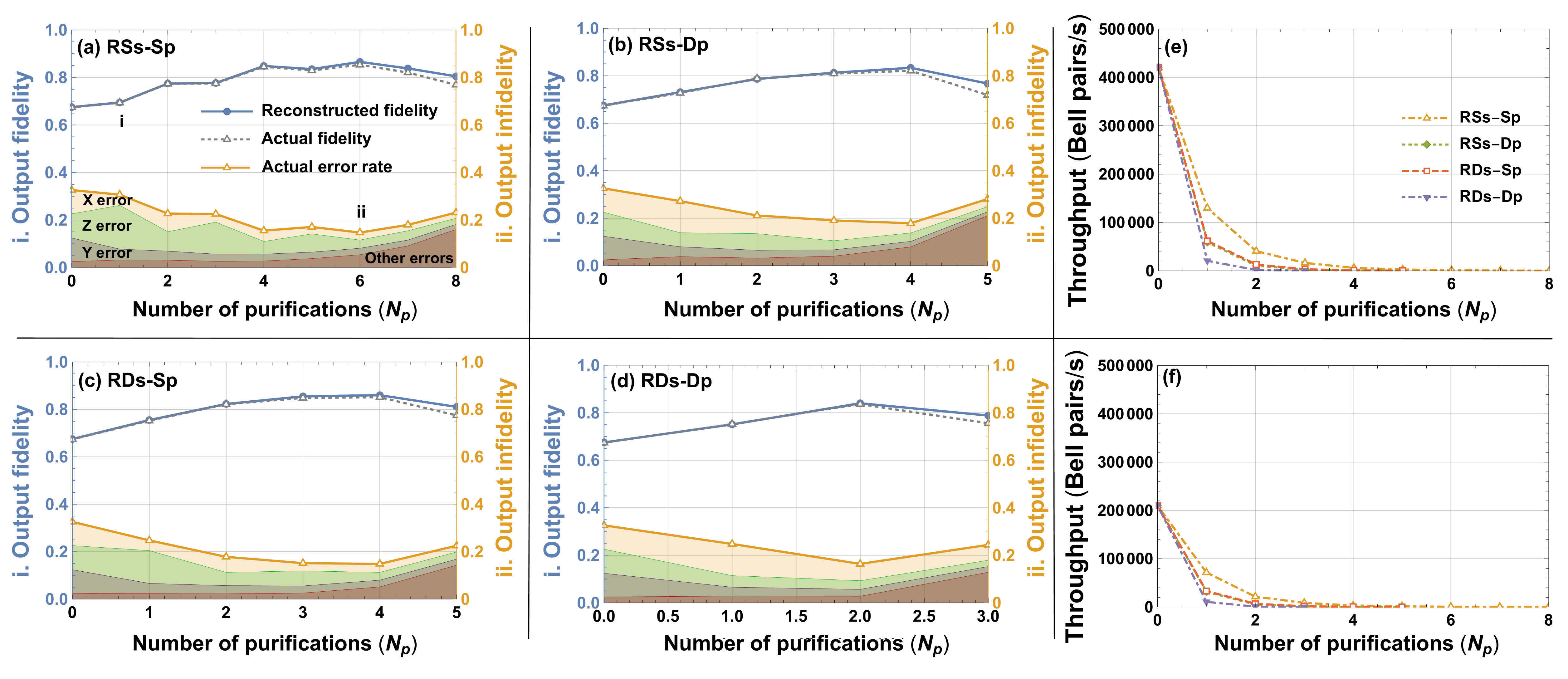}
   \caption{
   Reconstructed fidelity with actual fidelity and actual error rates, and estimated throughput of protocols over $L=10km$ MeetInTheMiddle link.
   "Other errors" include memory excitation/relaxation error, and completely mixed error due to photon detector dark counts.
   (a) Simulation result of the RSs-Sp protocol.
   (b) Simulation result of the RSs-Dp protocol.
   (c) Simulation result of the RDs-Sp protocol.
   (d) Simulation result of the RDs-Dp protocol.
   (e) Protocol throughputs over the MeetInTheMiddle link.
   (e) Protocol throughputs over the SenderReceiver link.
   }
   \label{nTimesPurification}
 \end{figure*}

We start by investigating how recurrent purification benefits over a link between closely located repeater nodes, $L=10km$.
The simulated result over the MeetInTheMiddle link is shown in Fig.~\ref{nTimesPurification}.
Notice that each protocol has a different number of purification rounds performed.

As in the figure, the given system generates resources with an average fidelity $F_{r}\approx 0.675$,
but the RDs-Dp is capable of bringing up the quality up to $F_{r}\approx 0.840$ through a two round purification,
which is the fastest in terms of required rounds.
The RSs-Sp, on the other hand, starts with the slightest fidelity improvement
since applying X purification inherently results in a significant penalty to the Z error rate, as we discussed in section \ref{purifications}.
While this looks operationally impractical, by alternating X and Z purification mulitple rounds,
we can purify resources while minimizing error propagations --
this appears as the stairstep-like curve in Fig.~\ref{nTimesPurification}(a).
RSS-Sp gradually improves the fidelity as we perform more purifications,
and produces resources with optimal average fidelity, $F_{r} \approx 0.865$ ($F_{r} \approx 0.852$ for the SenderReceiver link) at $N_p = 6$.
About 3.1\% of the errors are X errors, 2.7\% are Y errors and 3.4\% are Z errors.
5.6\% of the outputs consists of unfixable errors,
which can mainly be avoided by improving the memory lifetime, shortening the distance between nodes
or adjusting the algorithm and the goal.
RDs-Sp is also capable of producing high fidelity resources, $F_{r} \approx 0.860$,
with more rounds of purifications than RDs-Dp but fewer than RSs-Sp.
The fidelity starts declining for all protocols in the end,
due to critical waiting time of memories relative to its own lifetime.
The reconstructed fidelity also gets overestimated,
because the randomness of measuring unentangled qubits due to errors,
stochastically contributes to the fidelity estimation.

The number of rounds does not necessarily determine how fast the protocol processes resources, especially with a limited number of qubits.
As shown in Fig.~\ref{nTimesPurification},
the RDs-Dp protocol over the MeetInTheMiddle link reaches its highest fidelity with a throughput of roughly 1565/s,
completing 7000 purified Bell pair measurements in around 37.1 seconds.
The RSs-Sp protocol produces 1106 resources per second with the local optimal fidelity,
but also achieves similar fidelity as RDs-Dp, $F_{r}\approx 0.848$,
at $N_p=4$, which obtains a higher generation rate of 5923/s.
The RDs-Sp protocol and the RSs-Dp protocol have a throughput of 816/s and 520/s respectively.
Although both protocols consume the same amount of resources per purification,
the RSs-Dp took longer in total because at each step, RSs-Dp faces higher error propagation probability,
resulting in greater purification failure rate.
As provided in Fig.~\ref{nTimesPurification}(e) and Fig.~\ref{nTimesPurification}(f), the throughput over the SenderReceiver link is approximately half of what the MeetInTheMiddle link achieves.
In this case, RSs-Dp shows no advantage over RDs-Sp.

\begin{figure}[!htbp]
   \center
   \includegraphics[keepaspectratio,scale=0.45]{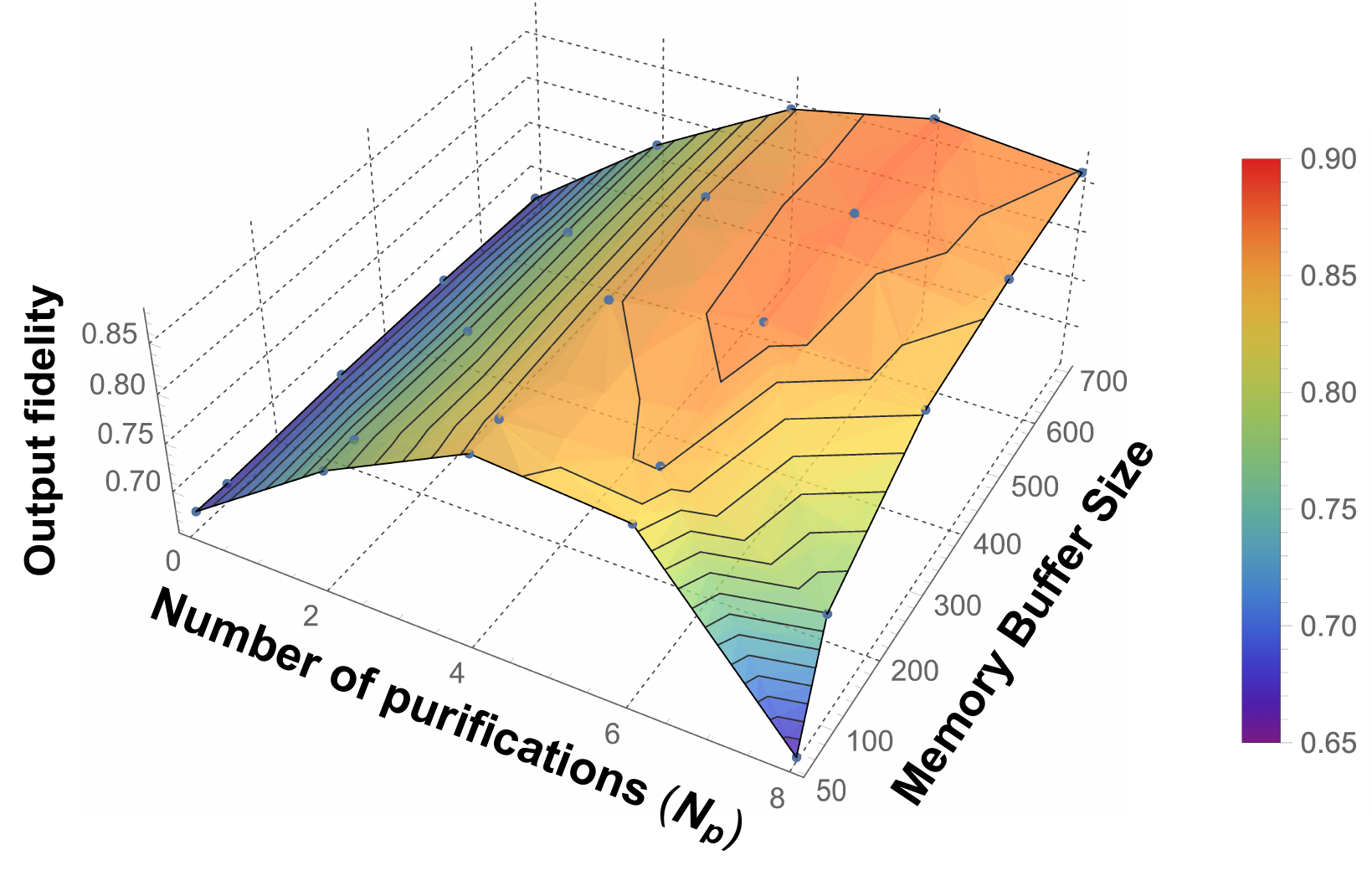}
   \caption{
   Impact of the memory buffer size to the maximum fidelity purified through the RSs-Sp protocol over a 10km MeetInTheMiddle link.
   }
   \label{memorysize}
 \end{figure}

The quality of a link can also be increased by installing larger sets of memory qubits.
As in Fig.~\ref{memorysize}, with 700 memory qubits, a six-round RSs-Sp can generate resources with $F_{r}\approx 0.879$.
Compared to the case with only 100 qubits, the fidelity at $N_{p}=4$ also increases around by 0.6\%.
For $N_{p}<4$, the fidelity stays unchanged because having 100 qubits is more than sufficient for a four-round RSs-Sp -- the throughput will, however, improve.
Notice that a linear increase in memory buffer size will not provide an effective solution to extend the performable rounds.
Over a 10km MeetInTheMiddle link, owning 700 memory qubits is still not enough for an eight-round RSs-Sp.

\subsubsection{Distantly spaced repeater nodes}

\begin{figure*}
  \center
  \includegraphics[keepaspectratio,scale=0.17]{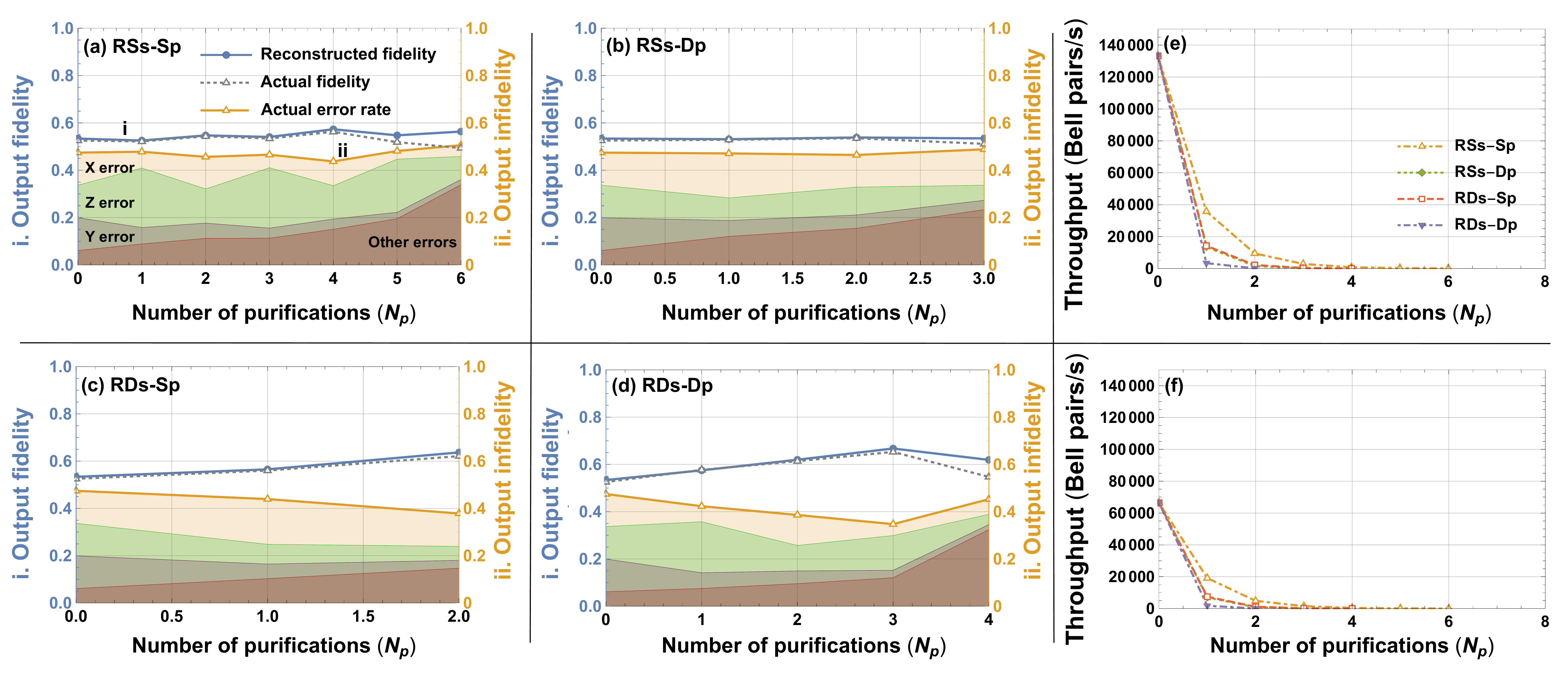}
  \caption{
  Reconstructed fidelity with actual fidelity and actual error rates, and estimated throughput of protocols over $L=20km$ MeetInTheMiddle link.
  "Other errors" include memory excitation/relaxation error, and completely mixed error due to photon detector dark counts.
  (a) Simulation result of the RSs-Sp protocol.
  (b) Simulation result of the RSs-Dp protocol.
  (c) Simulation result of the RDs-Sp protocol.
  (d) Simulation result of the RDs-Dp protocol.
  (e) Protocol throughputs over the MeetInTheMiddle link.
  (e) Protocol throughputs over the SenderReceiver link.
  }
  \label{nTimesPurification20km}
\end{figure*}

Errors on qubits have more time to develop when two repeater nodes are located farther away.
Hence, how well errors are detected at each round, especially with noisy inputs, becomes an important factor for an effective recurrence purification protocol.
The simulation results over a repeater network with channel length $L=20km$ are is provided in Fig.~\ref{nTimesPurification20km}.

Unlike the case of Fig.~\ref{nTimesPurification},
neither RSs-Sp nor RSs-Dp improves the fidelity effectively,
regardless of the number of purification rounds.
As shown in the corresponding error distributions, performing purifications with the given input resources,
in such a way done by RSs-Sp and RSs-Dp, ends up failing because the loss lead by error propagations and memory errors is greater than the gain.
Resources used for RSs-Sp suffer more from memory errors as they go through more rounds of purifications, resulting in longer average idle time per resource.
The other two protocols implemented with double selection,
however, are capable of effectively purifying resources.
The RDs-Sp protocol produces the highest fidelity among all,
with an output value $F_{r} \approx 0.667$ and a rate of 325/s (168/s with $F_{r} \approx 0.651$ for the SenderReceiver link) at $N_{p}=3$,
whereas RDs-Dp also produces similar quality, $F_{r} \approx 0.637$, or $F_{r} \approx 0.626$ for the SenderReceiver link, within two rounds.
The corresponding throughput is 116/s for the MeetInTheMiddle link, and 60/s for the SenderReceiver link.

Under this scenario, consuming another Bell pair via the double selection shows advantage in terms of fidelity.
However, from the previous discussion, we know that RSs-Sp generates resources with higher fidelity and higher rate,
given slightly better average input resources and sufficiently long-lived memories relative to its average idle time.
Thus, for longer distances, we can also perform double selection-based purification at the beginning to raise the fidelity at a certain level,
then change to single selection purification afterwards to maximize the fidelity.
Below in Fig.~\ref{MX20km} are the simulation results of two cases;
{\emph Case A} that switches to RSs-Sp after performing a single round of RDs-Sp,
and {\emph Case B} that switches to RSs-Sp after performing two rounds of RDs-Sp.

\begin{figure}[!htbp]
  \center
  \includegraphics[keepaspectratio,scale=0.23]{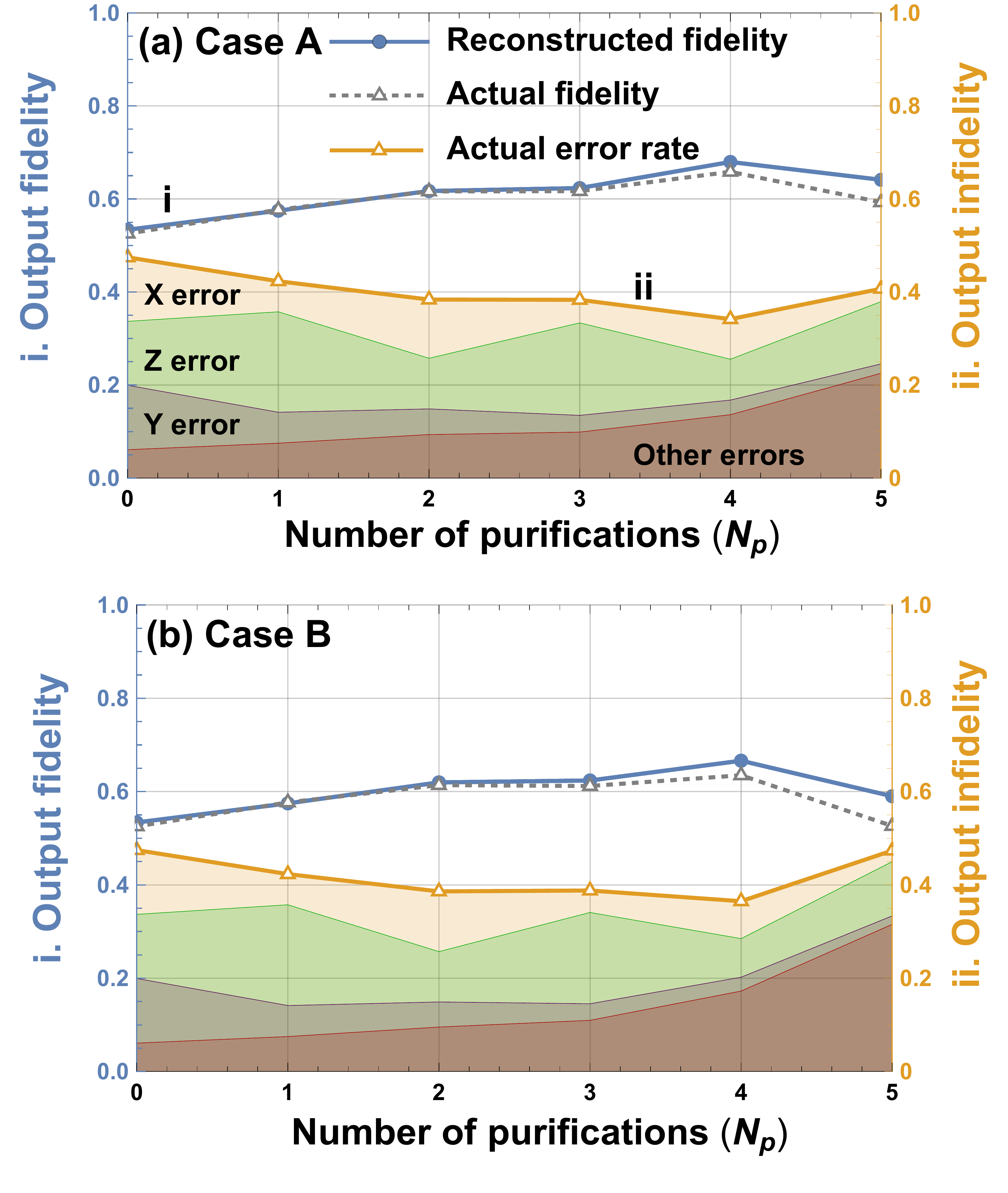}
  \caption{Switching RDs-Sp to RSs-Sp over $L=20km$.
  a) Pattern 1 consists of one round RDs-Sp and four rounds of RSs-Sp. Purification methodology switches at $N_{p}=1$.
  b) Pattern 2 consists of two rounds RDs-Sp and three rounds of RSs-Sp. Purification methodology switches at $N_{p}=2$.}
  \label{MX20km}
\end{figure}

Whether this method is beneficial or not very much depends on the situation, and the switching timing as shown.
Under this scenario, {\emph Case A} produces higher fidelity, $F_{r}=0.680$, with a higher rate of 396 Bell pairs per second,
relative to the optimal case of RDs-Sp in Fig.~\ref{nTimesPurification20km}.
{\emph Case B} results in worsening the fidelity to $F_{r}=0.666$, and the throughput to 218/s.

\section{Conclusion}

In this paper, we introduced our RuleSet-based communication protocol,
 which we use to autonomously support coordinated decision making in quantum operations over a network.
We adapted the technology to conduct quantum link bootstrapping in a network simulation.
In total, we ran four different recurrent purification schemes
to quantify the achievable link fidelity with its corresponding throughput via link-level tomography,
 using Markov-Chain Monte-Carlo simulation with a noisy quantum repeater system.
Such information may be distributed across the network, and used for different purposes such as quantum routing.

The RSs-Sp requires more purification rounds to optimize the fidelity compared to other schemes,
which results in a longer average idle time for purified resources.
Given 100 link memory qubits each with a 50ms lifetime,
we found that RSs-Sp is still stronger than other simulated schemes in terms of the achievable fidelity and the throughput,
when the distance between nodes is kept short ($L \approx 10km$).
In contrast, because the system is noisier with longer channels ($L \approx 20km$), RSs-Sp alone may be incapable of purifying the link,
 as errors can evolve faster than the purification, in which case schemes such as the double selection becomes more preferable.
Our simulations show that the bootstrapping process, therefore, must be able to search and choose the most effective purification method for that specific channel.
Changing from double selection to single selection in the middle indeed pushed the link limit,
but finding the optimal changeover timing may add an additional complexity to the process.
Other methods that detect errors with a higher probability requiring fewer rounds,
 such as RDs-Dp, may be more advantageous when a larger set of memory qubits is available for each link, relative to its demand.

While this work focuses on link fidelity optimization,
connecting source to destination multi-hops away also requires an adequately high generation rate, or possibly a synchronized generation timing of Bell pairs.
Such an optimization problem of the fidelity-throughput tradeoff remains an open question.

\begin{acknowledgments}
We thank Takahiko Satoh for valuable discussions.
This research was supported by the Q-LEAP program of Japan Science and Technology Agency (JST).

\end{acknowledgments}

\bibliography{LinkLevelTomography}

\appendix







\section{Memory error simulation}
\label{memoryerror}

We use a row vector to describe the present state of a qubit, which is one of the seven distinguishable states -- no error, X error, Z error, Y error, excited, relaxed or completely mixed.
For example, a Bell pair with no error as an initial input state can be described as in Eq.~\ref{inputvector}.
\begin{widetext}
\begin{eqnarray}
\label{inputvector}
\vec{\pi}(0) =\;\begin{blockarray}{cccccccc}
\scriptstyle Clean & \scriptstyle X{ }error & \scriptstyle Z{ }error & \scriptstyle Y{ }error & \scriptstyle Excited & \scriptstyle Relaxed & \scriptstyle Mixed \\
\begin{block}{(ccccccc)c}
  1 & 0 & 0 & 0 & 0 & 0 & 0 &  \\
\end{block}
\end{blockarray}
\end{eqnarray}
\end{widetext}

Accordingly, the infinitesimal generator for our continuous-time Markov-Chain~\cite{kleinrock1974qsv1} we use for the memory error simulation also consists of seven states,

\begin{widetext}
\begin{eqnarray}
\label{infinitesimalgenerator}
\overline{Q} =\;\begin{blockarray}{cccccccc}
\scriptstyle Clean & \scriptstyle X{ }error & \scriptstyle Z{ }error & \scriptstyle Y{ }error & \scriptstyle Excited & \scriptstyle Relaxed & \scriptstyle Mixed \\
\begin{block}{(ccccccc)c}
  (1-\Sigma_{0}) & P_X & P_Z & P_Y & P_E & P_R & 0 &  \\
  P_X & (1-\Sigma_{1}) & P_Y & P_Z & P_E & P_R & 0 & \\
  P_Z & P_Y & (1-\Sigma_{2}) & P_X & P_E & P_R & 0 & \\
  P_Y & P_Z & P_X & (1-\Sigma_{3}) & P_E & P_R & 0 & \\
  0 & 0 & 0 & 0 & (1-\Sigma_{4}) & P_R & 0 & \\
  0 & 0 & 0 & 0 & P_E & (1-\Sigma_{5})  & 0 & \\
  0 & 0 & 0 & 0 & P_E & P_R & (1-\Sigma_{6}) & \\
\end{block}
\end{blockarray}.
\end{eqnarray}
\end{widetext}

Element $P_X$, $P_Y$ and $P_Z$ correspond to Pauli X, Y and Z error accordingly.
$P_E$ and $P_R$ are memory excitation and relaxation errors,
which are used to represent $T_1$ time and the Boltzmann thermal distribution.
$\Sigma_{i}$ is the sum of every other element in row $i$.
Mixed state occurs due to dark counts, or when the paired qubit gets excited/relaxed.
The probability distribution of a qubit output state after time $t$ can be found by:

\begin{eqnarray}
 \vec{\pi}(t)=\vec{\pi}(0)\overline{Q}^t.
\end{eqnarray}

Each element in the output vector $\vec{\pi}(t)$ represents the probability for the qubit being in the corresponding state.
The qubit state can be determined via a random selection based on the output probability distribution,
 and the density matrix is constructed on demand.

\end{document}